%% file: main.tex
\documentclass[a4paper,11pt]{article}
\usepackage{jinstpub} 
\usepackage{lineno}
\usepackage{float}
\usepackage{graphicx} 

\pdfoutput=1
\usepackage[english]{babel}
\usepackage{upgreek}
\usepackage{xspace}
\usepackage[capitalize]{cleveref}
\usepackage{enumitem}

\title{\boldmath Long-term calibration and validation of 
 stability of the Auger Engineering Radio Array using the diffuse Galactic radio emission}

\author{\includegraphics[height=30mm]{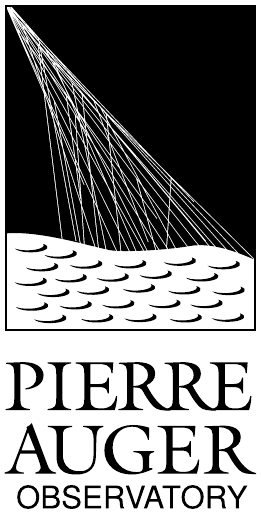}\\[3mm]The Pierre Auger Collaboration}
\affiliation{Av.\ San Mart\'{\i}n Norte 306, 5613 Malarg\"ue, Mendoza, Argentina}

\emailAdd{spokespersons@auger.org}

\abstract{The Auger Engineering Radio Array (AERA) measures radio emission from high-energy extensive air showers. Consisting of 153 autonomous radio-detector stations spread over $17$\,km$^2$, it detects radio waves in the frequency range of $30$ to $80$\,MHz. Accurate characterization of the detector response is crucial for proper interpretation of the collected data. Previously, this was achieved through laboratory measurements of the analog chain and simulations and measurements of the antenna's directional response. In this paper, we perform an absolute calibration using the continuously monitored sidereal modulation of the diffuse Galactic radio emission. Calibration is done by comparing the average frequency spectra recorded by the stations with predictions from seven different models of the full radio sky, accounting for the system response, which includes the antenna, filters, and amplifiers. The analysis of the calibration constants over a period of seven years shows no relevant and no significant ageing effect in the AERA antennas. This result confirms the long-term stability of the detector stations and demonstrates the possibility for a radio detector to effectively monitor ageing effects of other detectors operating over extended periods. }


\begin{document}
\maketitle
\flushbottom

\section{Introduction}
\label{sec:intro}
\hspace{0.5cm} Since the discovery of extensive air showers in the late 1930s, the origin of high-energy cosmic rays has been a mystery~\cite{CosmicRay1,CosmicRay2,CosmicRay3,CosmicRay5}. In particular, for ultra-high-energy cosmic rays (UHECRs), there are still significant gaps in our understanding regarding their sources, mass composition, acceleration mechanisms, and propagation to Earth.

The Pierre Auger Observatory~\cite{Auger1} is situated near Malargüe, in the province of Mendoza, Argentina, and spans $3000$ km$^2$, the largest observatory dedicated to the study of UHECRs. Its hybrid design combines a surface detector (SD) array of 1660 water-Cherenkov stations with a fluorescence detector (FD) comprising 27 telescopes at four sites that overlook the array~\cite{Auger1,FDPaper}. Within the SD grid, the Auger Engineering Radio Array (AERA)~\cite{AERA} provides complementary radio measurements, operating in the frequency range $30-80$\,MHz with two channels, each one measuring in a different polarization: east-west and north-south relative to magnetic North. AERA is currently the second largest system designed to measure radio emissions from ultra-high-energy extensive air showers, only surpassed by the recently deployed $3000$\ km$^2$ Radio Detector~\cite{ref:RDICRC2023}, also at the Pierre Auger Observatory. 
Illustrative of the advancements made by radio detectors, particularly AERA, are their capabilities in estimating the energy of cosmic rays~\cite{AERA_energy,AERA_energy2} and using radio measurements to determine the depth of shower maximum ($X_{\text{max}}$) in air showers induced by cosmic rays~\cite{XmaxRadio1,XmaxRadio2}.

A precise characterization of the detector's response is crucial for the accurate interpretation of the data collected by the stations. In the past, this was accomplished by conducting measurements on the analog chain in the laboratory, as well as simulating and measuring the directional response of the antennas using a drone~\cite{Calibration_LPDA}. This type of calibration, utilizing a reference antenna emitting a predetermined signal, has the drawback of uncertainties regarding the emitted signal strength~\cite{lofar}. Moreover, conducting dedicated calibration campaigns requires significant effort and is nearly impractical to execute regularly, which is especially crucial when evaluating the long-term performance of the detector.  These long-term effects, commonly referred to as detector ageing, arise from the natural degradation of detector components over time and can gradually affect the quality and stability of the recorded data. Monitoring and correcting for such effects are standard practices at the Pierre Auger Observatory, particularly for the Surface Detector Array and the Fluorescence Telescopes~\cite{PerformanceSD,PerformanceFD}.

In this study, we adopt a different approach by performing an absolute Galactic calibration of the AERA antennas.  Calibration constants are obtained by comparing the average spectra recorded by the stations with a comprehensive model of the entire radio sky propagated through the system response, taking into account the antenna, filters, and amplifiers. Despite the fact that we commonly designate it as Galactic emission in this study, it implicitly considers minor extragalactic components.  Additionally, we study the behavior of the calibration constants over time from 2014 until 2020. The results show that no significant ageing effect has been observed in the AERA antennas over almost a decade, highlighting the potential of radio detectors not only for precise data interpretation but also for monitoring ageing effects in other detectors during long-term operation.

The paper is structured as follows: Section~\ref{sec:data_set} describes the detector system and the data set used for the calibration process. Section~\ref{sec:model} describes the radio sky models employed in this study, while Section~\ref{sec:method} presents a comprehensive explanation of the methodology adopted for the absolute Galactic calibration, as well as the results. The investigation of calibration constants over time is discussed in Section~\ref{sec:c0_time}. Finally, Section~\ref{sec:conclusions} summarizes the conclusions drawn from the study.

\section{Detector system and data set}
\label{sec:data_set}
\hspace{0.5cm} 
Since AERA is an engineering array, various antenna types, electronics, and trigger systems have been developed, deployed, and tested in the field over time.  The deployment occurred in three phases and the distribution of the electronics across the antenna stations changed multiple times. In this study, we use data measured by butterfly antennas and Logarithmic-Periodic Dipole Antennas (LPDA)\cite{AERA} at detector stations that can handle external triggers provided by the baseline Auger detectors. LPDAs use the logarithmic periodic dipole principle, incorporating a series of half-wave dipoles with progressively increasing lengths to ensure consistent radiation resistance across a broad frequency spectrum~\cite{Stations}. 
In contrast, butterfly antennas, commonly referred to as "bow-tie" models, have a much simpler mechanical design, consisting of two triangular arms per polarization.

The absolute Galactic calibration is performed by using periodically triggered traces measured with 52 butterfly antennas from 2014 until 2020, 23 butterfly antennas with data collected from January 2016 until 2020, and 14 LPDAs with available data from 2017 to 2020. Periodically triggered traces refer to the read-out requests made by the data acquisition system for all active stations every 100 seconds. The traces are corrected for an observed anti-correlation between peak amplitude and temperature caused by temperature-dependent gain variations of the Low Noise Amplifier (LNA) and Filter Amplifier (FA) in the signal chain, which had been characterized previously in lab measurements~\cite{Temperature}. Unlike for the FA, there is no temperature sensor located near the LNA. Given that the LNA is placed on top of the antenna pole, its temperature is assumed to be the local ambient temperature recorded at the Central Radio Station (CRS), which, among other AERA services, hosts a meteorological station.

The time series of each trace is clipped to 1024 samples ($5.7$\,\textmu s), which corresponds to a frequency spectrum of 285 bins, with a width of $\Delta \nu = 0.175$\,MHz, within the range of $30-80$\,MHz. For Galactic calibration purposes, we compute the power within each frequency band $\nu$, which has a width of $\delta \nu = 1$\,MHz, by using

\begin{equation}
\label{eq:power}
    P_\nu = \frac{2}{t}\sum_{k=\nu - \delta \nu/2}^{\nu+\delta \nu/2}\frac{|V(k)|^2}{Z_L} \Delta \nu,
\end{equation}
where $t$ is the length of the trace, $V(k)$ is the  measured spectral voltage at frequency $k$ and $Z_L= 50$\,$\Omega$ is the antenna impedance. The factor of 2 arises from utilizing only half of the Fast Fourier Transform spectrum. Since all stations are located in approximately the same geographic region, the Galactic signal can be assumed to be the same for all antennas. Nevertheless, the background signal measurement encompasses a combination of the Galactic component and various other types of noise. The intensity of these noise components can vary depending on the type and distance of the signal sources. Hence, it is crucial to preprocess the data by mitigating these noise components before proceeding with the Galactic calibration.

\subsection{Preprocessing of data}
\label{subsec:noises}

\begin{figure}[H]
        \centering
                \includegraphics[scale=0.17]{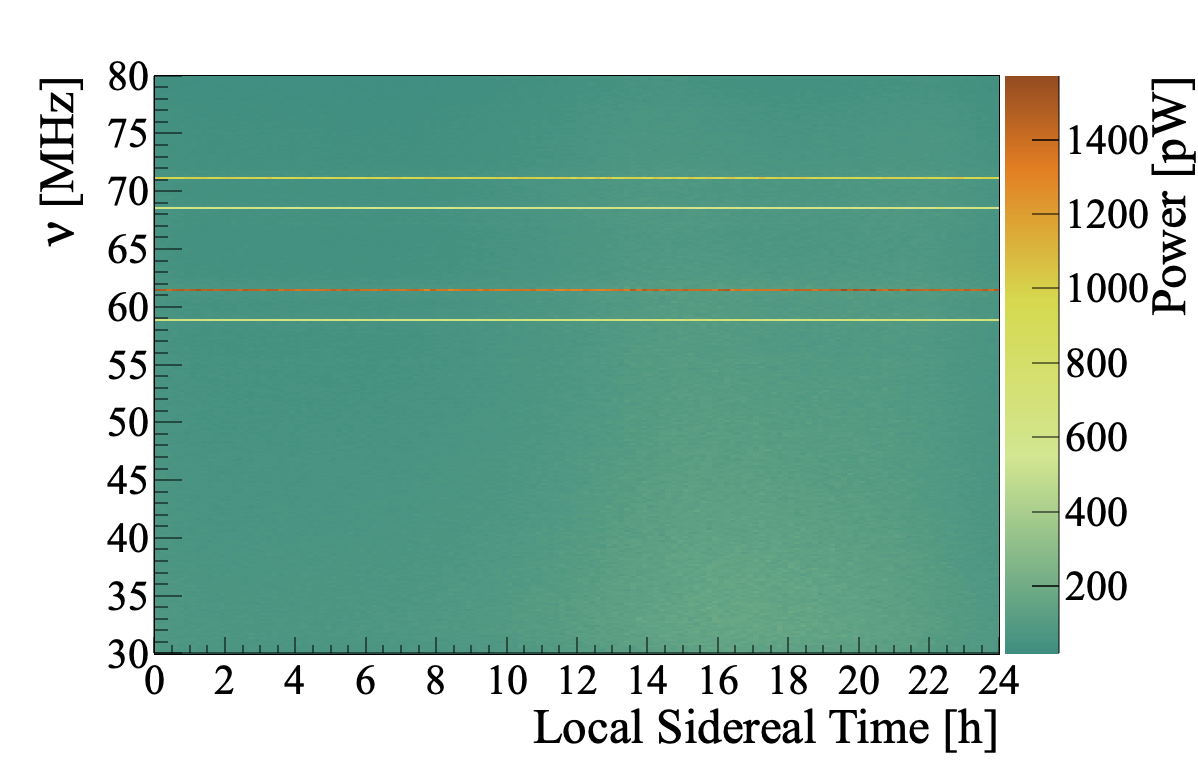} \quad
                \includegraphics[scale=0.17]{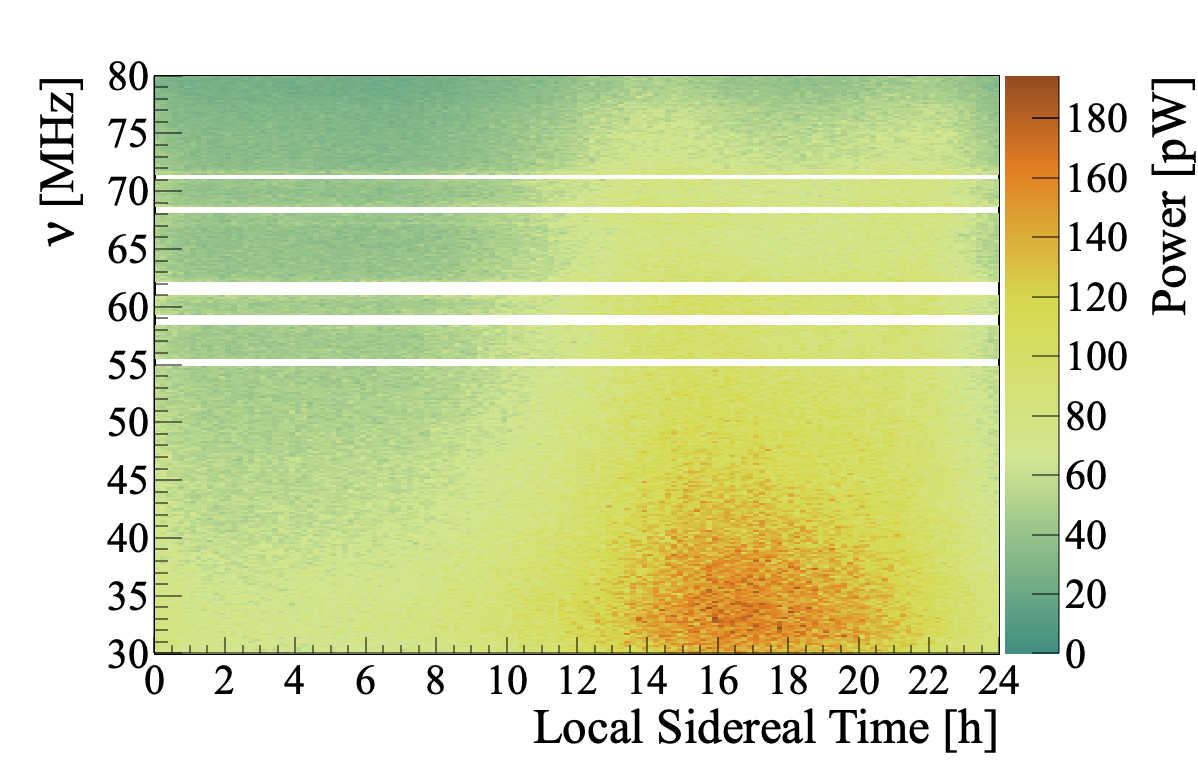} 
        \caption{\footnotesize{Dynamic average frequency spectrum as a function of LST for the north-south channel of one of the antennas during February 2019. The left panel presents the results before removal of narrowband RFI (horizontal lines), while the right panel shows that the Galactic signal variation becomes evident after RFI removal.}}
                \label{Fig:Spectrum_EW_NS}
 \end{figure}

\hspace{0.5cm} A Galactic modulation of the radio signal intensity as a function of Local Sidereal Time (LST) results from the passage of the Galaxy above the Auger site.  This modulation is observed by using periodic data. However, the data also contain cosmic ray signals, radio frequency interference (RFI) from external sources, and internal electronic noise. RFIs can be classified into two categories: broadband and narrowband. Narrowband noise is characterized by continuous emissions in the entire local sidereal time range and is produced by sources that continuously emit radio signals at specific frequencies. The most intense narrowband noise contributions are produced by the AERA beacons transmitter ($58.9$\,MHz, $61.5$\,MHz, $68.5$\,MHz, $71.2$\,MHz) installed at the Auger site for time calibration purposes~\cite{beacon}. Other narrowband RFI is observed at $67$\,MHz, due to a TV line, and at $55$\,MHz, corresponding to an unverified source that coincides with known TV line frequencies. The amplitude of these noise contributions makes it difficult to detect the passage of the Galaxy center in the dynamic spectrum. To overcome these issues, these frequency bands are identified and then replaced with values calculated by interpolating across the corresponding gaps. Figure~\ref{Fig:Spectrum_EW_NS} illustrates the average frequency dynamic spectrum obtained from one of the antennas during February 2019 before and after RFI removal for the north-south channel. One can see that the Galactic modulation is highlighted after the noise suppression. 

On the other hand, broadband RFIs are transient radio pulses that are also sources of contamination in the data. To mitigate the effects of broadband noise and cosmic ray signals, we implemented a threshold to discard undesired traces as a function of LST. For this, we first compute the average spectral density of each trace according to

\begin{equation}
\label{eq:averageSpectralDensity}
     I = \frac{1}{n}\sqrt{\sum_{i=1}^{n}A^{2}(\nu_i)},
\end{equation}where $A(\nu_i)$ is the signal amplitude in frequency bin $i$ and $n$ is the total number of bins. The time-dependent threshold is determined in the following way. For each 20-minute bin in LST, a Gaussian fit is performed on the distribution of $I$, and a threshold $I^{\mathrm{th}}_{\mathrm{LST}}$ corresponding to 3$\sigma$ is obtained by solving $\int_{-\infty}^{I^{\mathrm{th}}_{\mathrm{LST}}} G(I;\bar{I},\sigma_I) = 0.9973$, where $\bar{I}$ and $\sigma_I$ are the average and standard deviation of the Gaussian distribution, respectively. Traces with average spectral densities $I$ greater than $I^{\mathrm{th}}_{\mathrm{LST}}$ are removed from the data to suppress these transient noise. Finally, a smoothing is applied to the threshold $I^{\mathrm{th}}_{\mathrm{LST}}$ in order to prevent large variations between time bins.

\begin{figure}[H]
        \centering
                \includegraphics[scale=0.36]{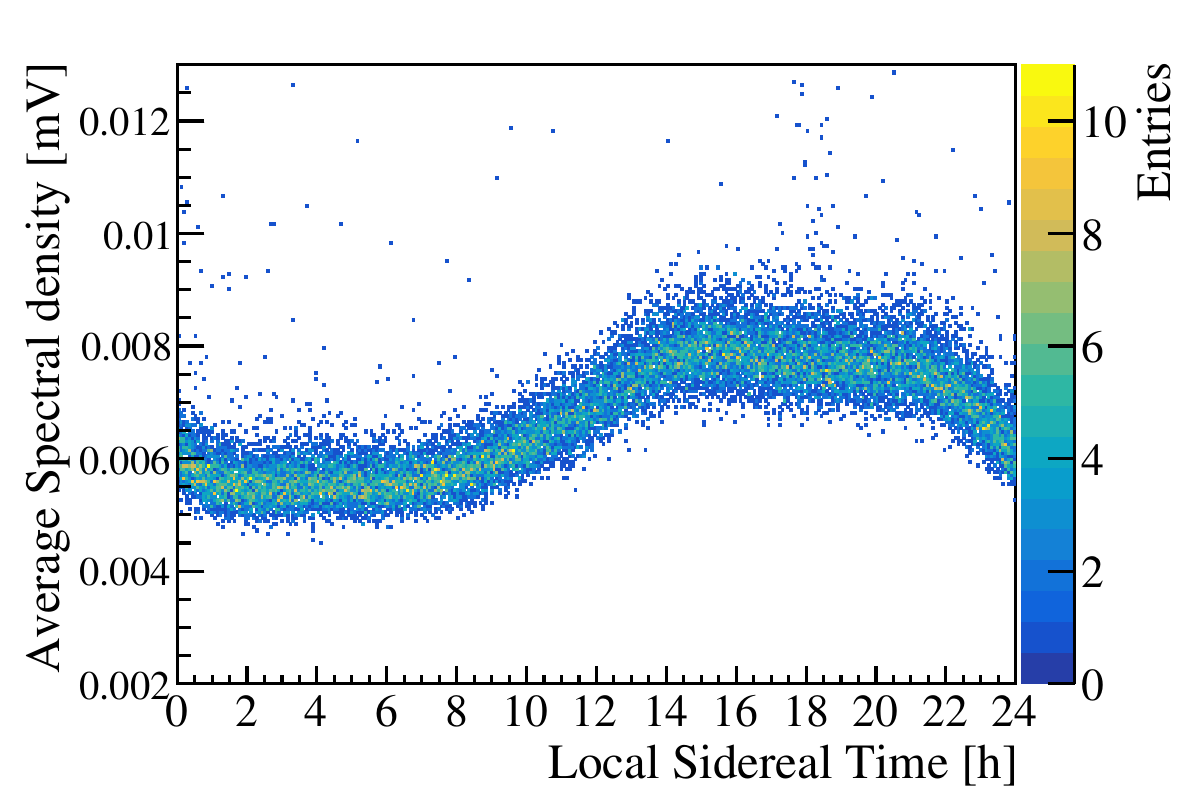} \quad
                \includegraphics[scale=0.36]{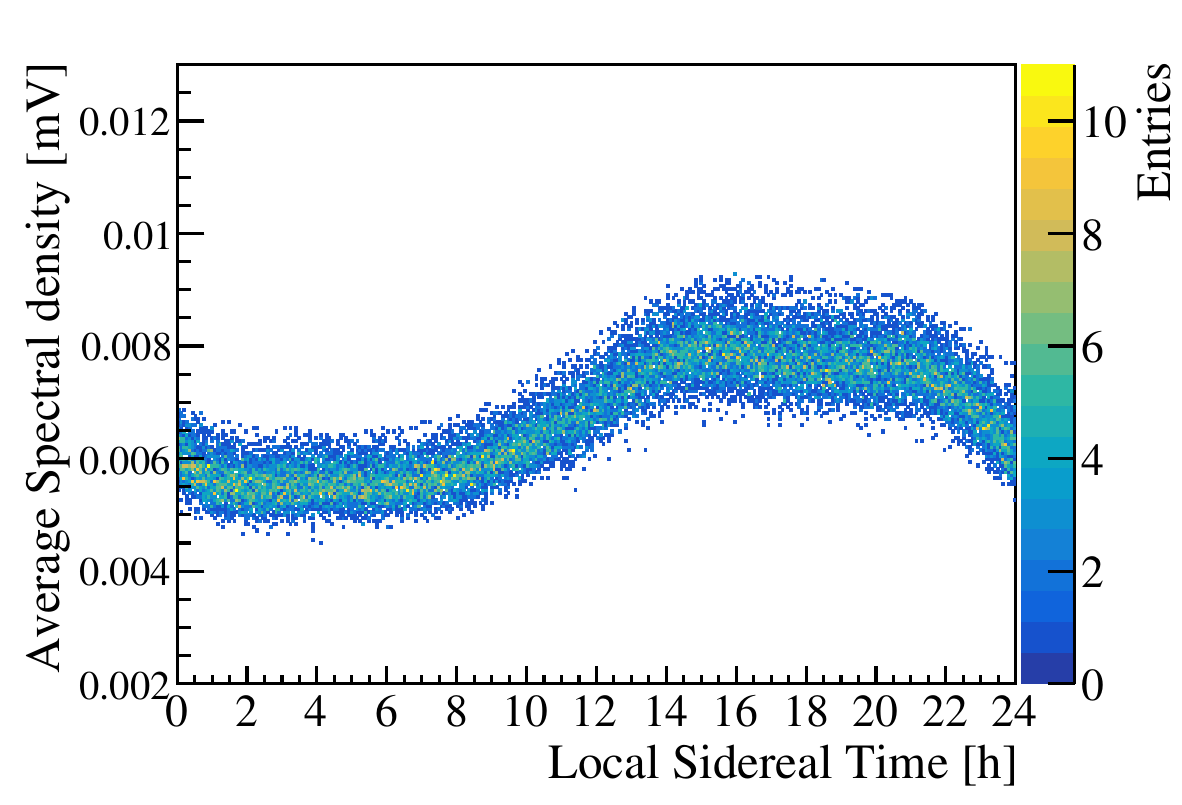} \quad
        \caption{\footnotesize{Average spectral density for one of the antennas during January 2019 as a function of LST for the north-south channel, before the removal of broadband RFI (left) and after the removal (right).}}
                \label{Fig:broadBand_spectral}
 \end{figure}

 Figure~\ref{Fig:broadBand_spectral} shows the distribution of the average spectral density for one of the stations during January 2019 as a function of LST, both before and after the removal of RFI. For the north–south channel of the Butterfly antennas, the typical Galactic signal is of the order of $0.0055$\,mV when the Galactic plane is outside the field of view (around LST $\sim$5\,h) and about $0.0075$\,mV when it is within the field of view (around LST $\sim$18\,h). The presence of highly intense broadband noise pulses is evident, reaching average spectral densities of more than $0.012$\,mV. After the RFI cleaning, significantly cleaner data are obtained, with average spectral density values below approximately $0.009$\,mV. 

\section{Radio sky models}
\label{sec:model}

\hspace{0.5cm} The background radio signal received on Earth varies in different directions across the sky and can be conveniently specified by its equivalent brightness temperature. In this way, radio maps of the sky are produced using measured brightness temperatures. Within the frequency range of the AERA stations, the background signal is dominated by Galactic emission, and the total expected power received by the antenna is calculated as
\begin{equation}\label{eq:model}
    P_{\mathrm{sky}}(t,\nu) = \frac{Z_0}{Z_\mathrm{L}}\frac{k_\mathrm{B}}{c^2} \int_{\Omega}\nu^2 T_{\mathrm{sky}}(t,\nu,\theta,\phi)|H(\nu,\theta,\phi)|^2 d\Omega,
\end{equation}where $k_\mathrm{B}$ is the Boltzmann constant, $c$ is the speed of light, $Z_{0}$ is the impedance of free space and $Z_\mathrm{L}$ is the antenna impedance. The integration is performed over the solid angle $\Omega$, corresponding to the region of the sky above the horizon that is visible at each local time $t$. $H(\nu,\theta,\phi)$ is the Vector Effective Length (VEL) of the antenna that is represented by a 2-dimensional vector with complex entries and gives the directional response as a function of the frequency of a radio signal coming from the direction $(\theta,\phi)$ given by~\cite{Stations}:

\begin{equation}\label{eq:directionalresponse}
\left | H(\nu,\theta,\phi) \right |^2 = H_{\theta}(\nu,\theta,\phi)^2 + H_{\phi}(\nu,\theta,\phi)^2.
\end{equation} $H_{\theta}$ is the complex response of the antenna towards signals being polarized in the $\theta$ direction, and $H_{\phi}$ the one in the $\phi$ direction.  The VEL has been simulated for both the LPDA and butterfly antenna types using a package specifically designed for simulating antenna characteristics~\cite{NEC2,Stations}. An in-situ measurement campaign was conducted in 2015 to ascertain the VEL of LPDA and butterfly antennas. Measurements were taken from various directions using a transmitting antenna and a signal generator mounted on an octocopter drone. The overall uncertainty in the measured VEL of the LPDA is $7.4\%$  for $H_{\phi}$ and $10.3\%$ for $H_{\theta}$ components~\cite{Calibration_LPDA}. For the Butterfly antennas, however, a discrepancy between the simulations and measurements was observed in the asymmetry of the response, which is mostly attributed to the electronic box mounted below the antenna. The Galactic background assessed using the periodically triggered traces from butterfly antennas is better described by the simulated directional response than by the measurements from the octocopter campaign. Therefore, the Galactic calibration reported in this work uses the simulated VEL for butterfly antennas and the measured VEL for LPDAs, as defined in Equation \ref{eq:directionalresponse}.

$T_{\mathrm{sky}}(t,\nu,\theta,\phi)$ represents the sky brightness temperature at frequency $\nu$, expressed in local coordinates as a function of the zenith angle $\theta$ and azimuth angle $\phi$, for each time $t$. There are different available models for the  sky radio emission such as the Low-Frequency Sky Map Generating Program (LFmap)~\cite{lfmap}, Global Sky Model (GSM 2008)~\cite{GSM}, (GSM 2016)~\cite{GSM16}, Low Frequency Sky Model (LFSM)~\cite{LFSM}, Global MOdel for the radio Sky Spectrum (GMOSS)~\cite{GMOSS},  Sky foreground Model (SSM)~\cite{SSM}, Ultra-Long-wavelength Sky Model with Absorption (ULSA)~\cite{ULSA} and others that will not be covered in this study. A detailed comparison and discussion of these models has been presented in \cite{busken2022}. Here, we only give a concise summary of their main characteristics.

The LFmap, GSM 2008, GSM 2016, and LFSM models employ different interpolation techniques based on reference sky maps to comprehensively characterize the radio-frequency sky within a specific frequency range. LFmap is a program designed to produce sky maps from tens to hundreds of MHz frequencies. This program scales the $408$\,MHz all sky map by Haslam~\cite{Haslam} to lower frequencies based on the simple power-law model for the sky brightness temperature. The GSM 2008 model employs Principal Component Analysis (PCA) to generate all-sky maps at any intermediate frequency. The GSM 2016 model is an updated version of the original GSM 2008. It considers further reference maps,  extends the frequency range, and enhances accuracy across all sub-GHz frequencies. As in~\cite{busken2022}, we consider both versions of the GSM model in this study because the original one is still widely used and most of the newly added reference maps of the updated version are at higher frequencies. The LFSM model, also PCA-based, utilizes sky surveys from the Long Wavelength Array Station 1 (LWA1) to generate radio maps in the frequency range of 10 to $408$\,MHz.

\begin{figure}[H]
        \centering
            \includegraphics[scale=0.366]{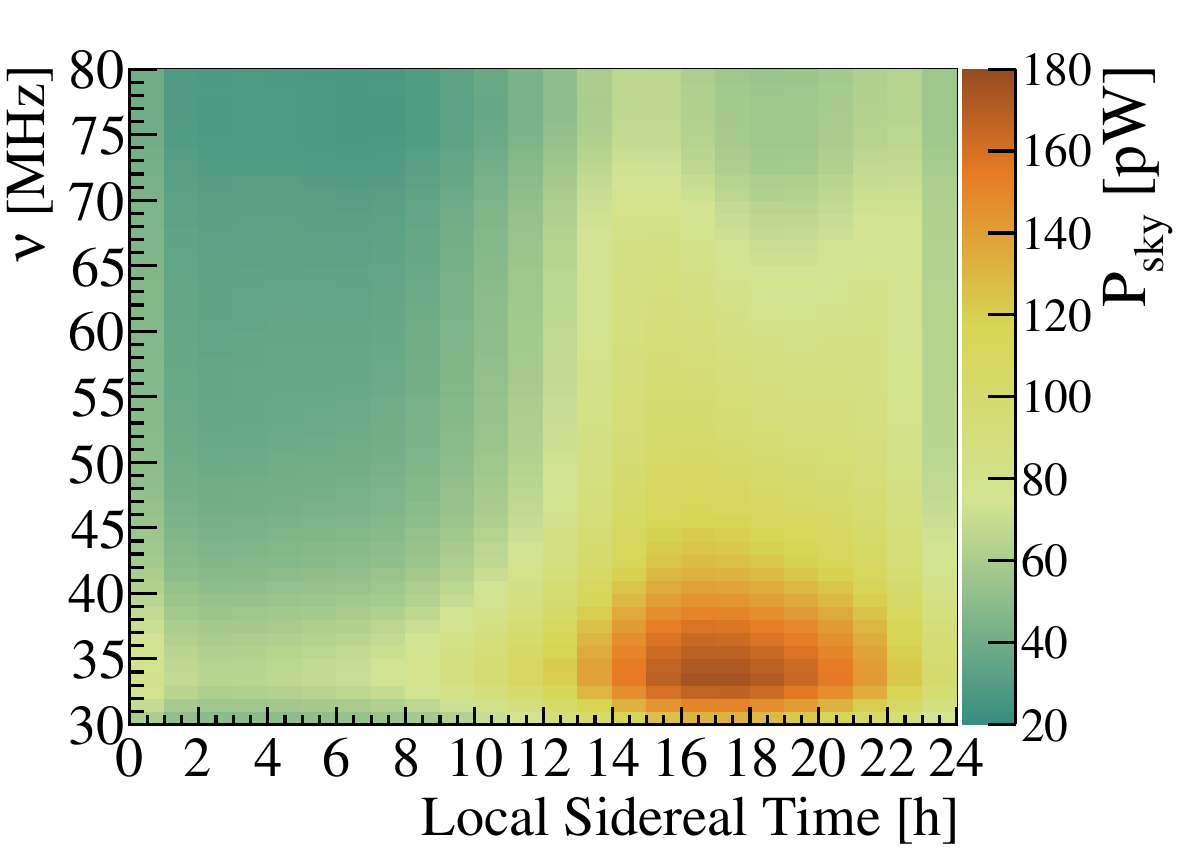} \quad
                \includegraphics[scale=0.366]{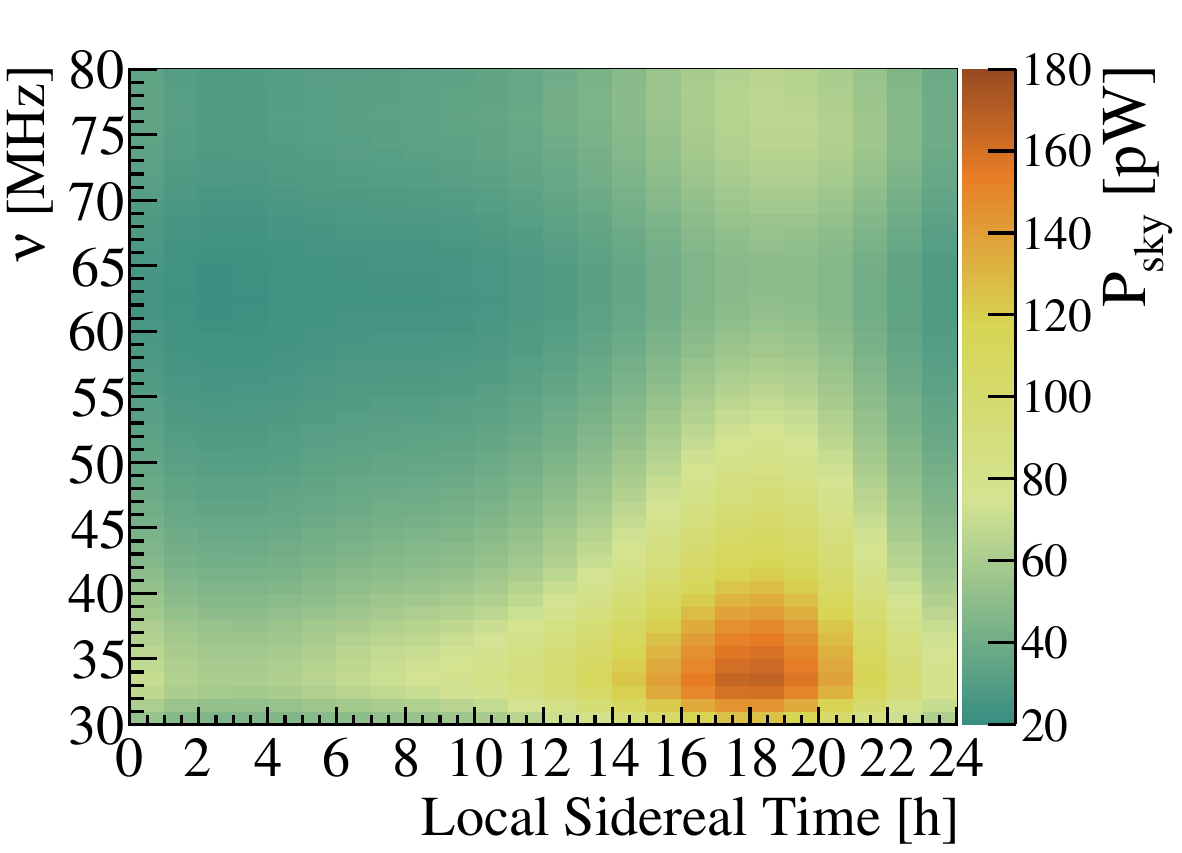} \quad
        \caption{\footnotesize{The left (right) panel shows the expected power $P_{\rm{sky}}$ to be received from the sky considering the LFmap model and the simulated VEL of butterfly antennas, in frequency bins of 1 MHz and LST bins of 1 hour, as a function of LST and frequency for the north–south (east–west) channel. }}
                \label{Fig:p_model}
 \end{figure}
 
In contrast, GMOSS is a physical model incorporating various mechanisms of diffuse radio emission, such as synchrotron radiation, thermal radiation, and free-free emission, to compose frequency maps of the sky. The SSM model estimates diffuse emissions using a conventional power-law spectrum and point sources using two different catalogs, extracting the spatial distribution of sources and the spectral index behavior. Finally, the ULSA model excels at ultra-long-wavelength radio emission and incorporates free-free absorption in the Galaxy below $10$\,MHz. In this model, the sky temperature is considered as the sum of a Galactic component and an isotropic extragalactic component, with a temperature reduction due to Galactic absorption. An estimate of the systematic uncertainties
in the prediction of the Galactic emission arising from the variations between these sky models was obtained in~\cite{busken2022}. As an illustration, the expected power $P_{\mathrm{sky}}(t,\nu)$ to be received from the sky by considering the LFmap model and the simulated VEL of butterfly antennas, in frequency bins of $1$\,MHz and LST bins of 1 hour, is shown in Figure~\ref{Fig:p_model} for both polarizations.

\begin{figure}[H]
        \centering
                \includegraphics[scale=0.36]{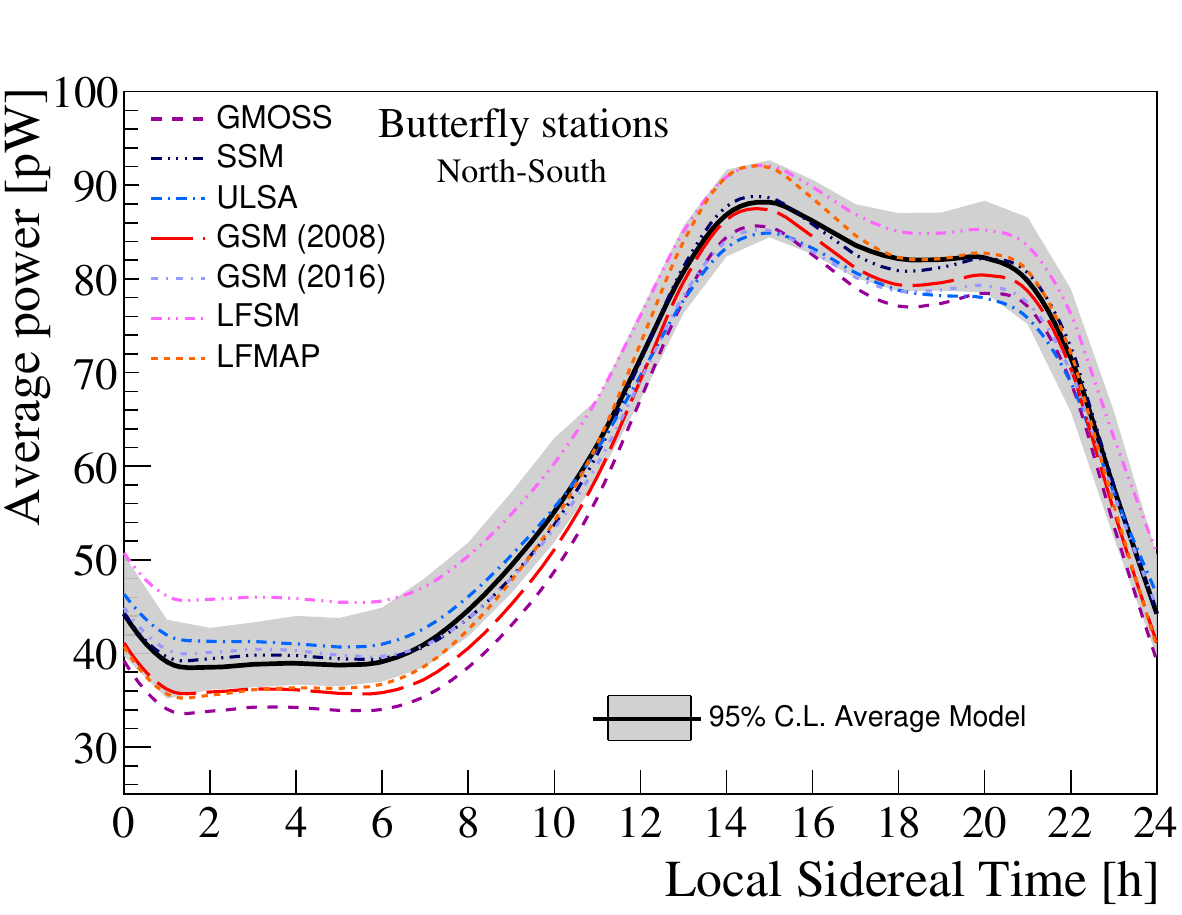} \quad
                \includegraphics[scale=0.36]{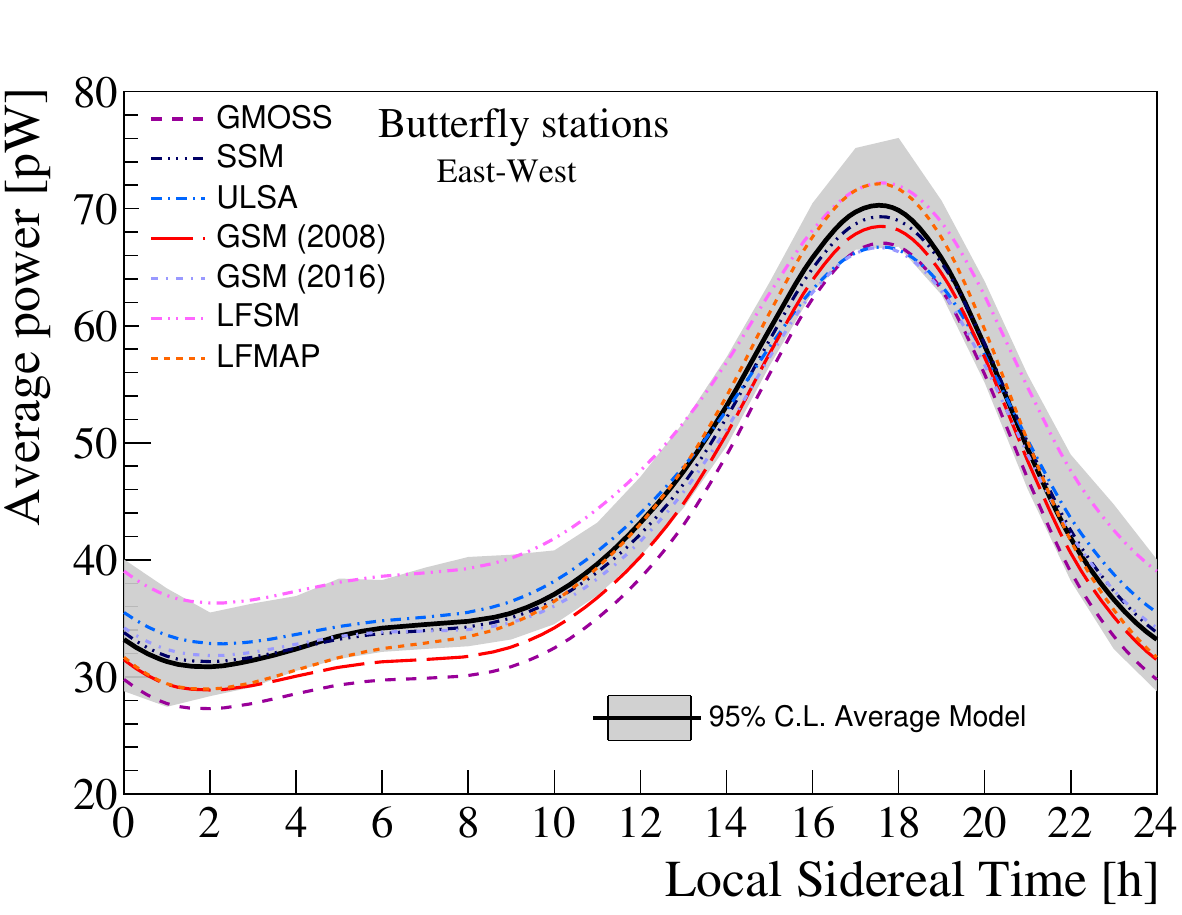} \quad
                \includegraphics[scale=0.36]{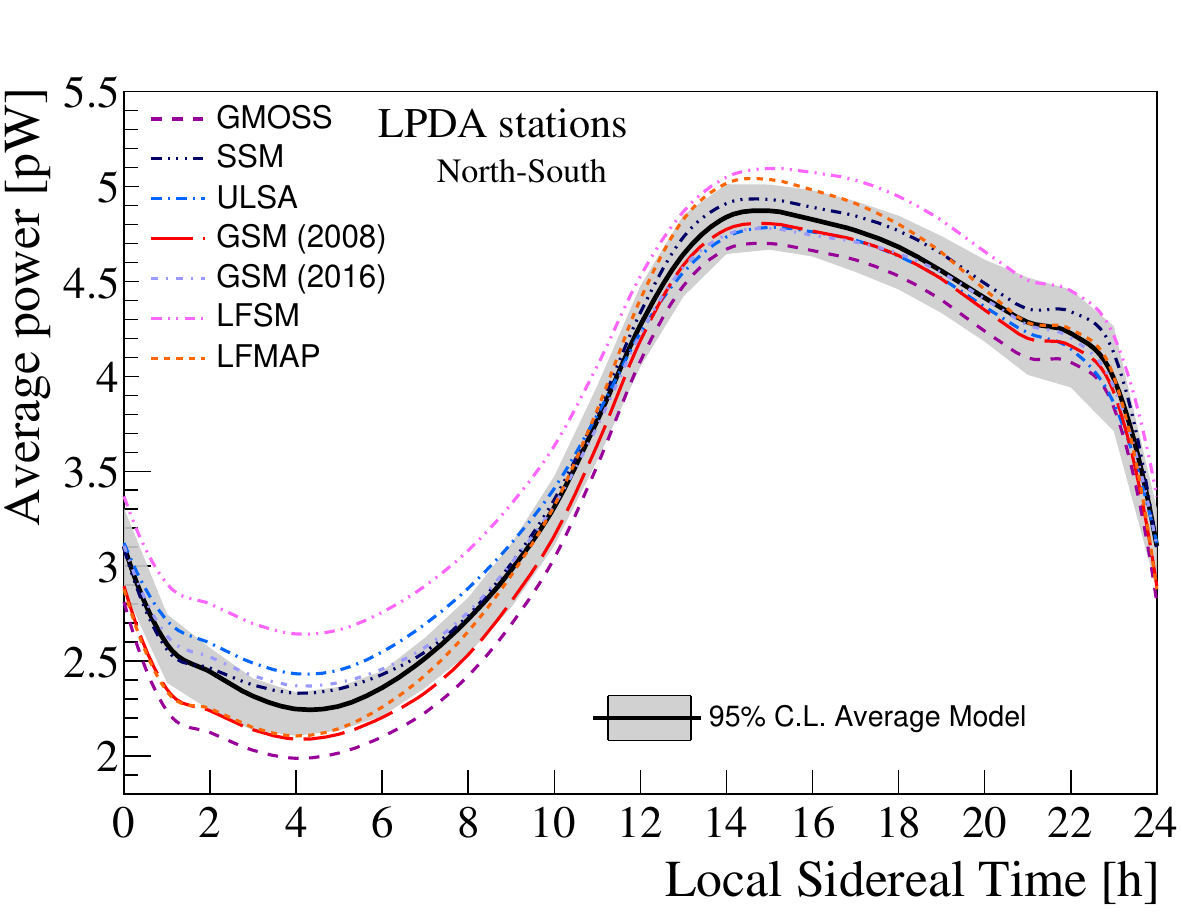} \quad
                \includegraphics[scale=0.36]{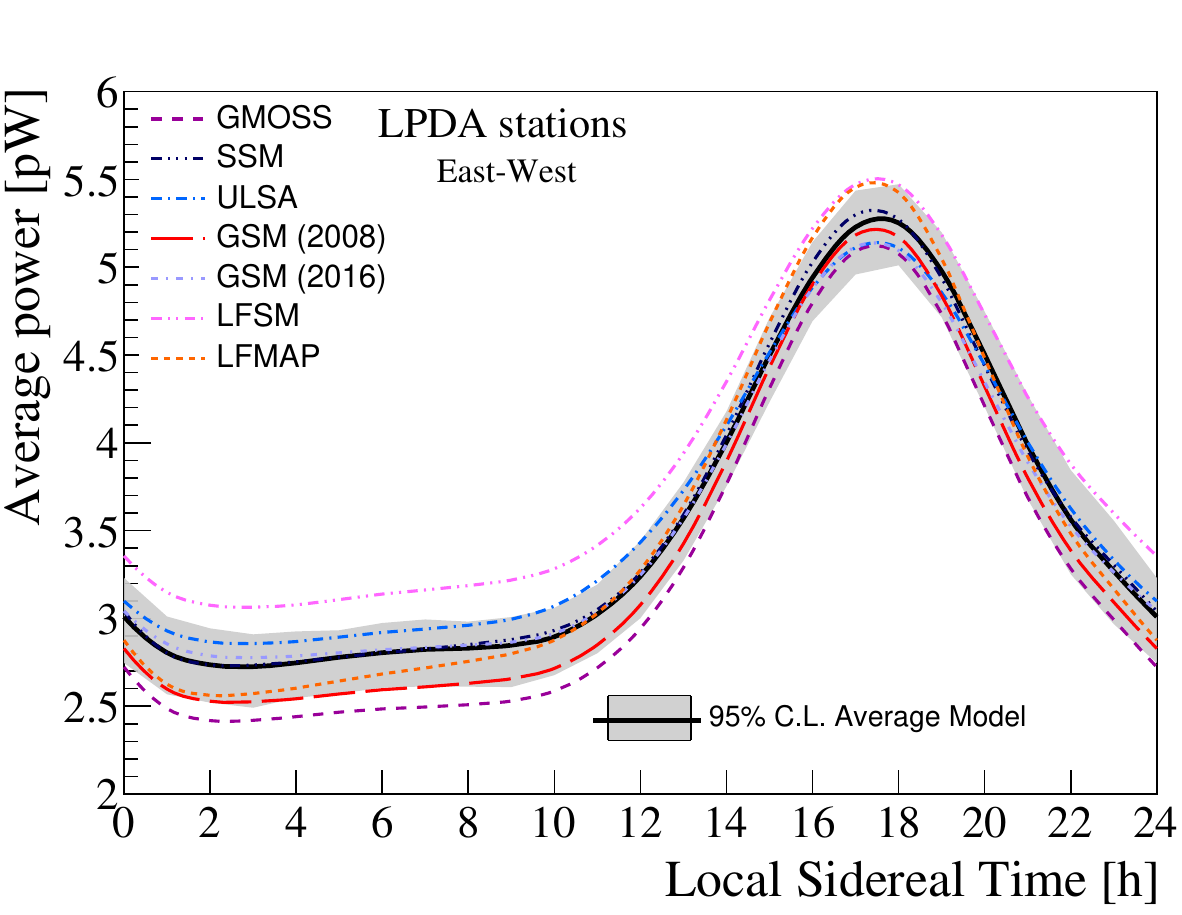} \quad
        \caption{\footnotesize{Averaged power over frequencies for each radio sky temperature model propagated through the antenna pattern as a function of time. Panels on the left and right represent the north-south and east-west polarizations, respectively. Panels above (below) correspond to butterfly antennas (LPDAs). The solid black line represents the average model, and the shaded areas indicate two standard deviations of the average power distribution calculated for each antenna after correction with the calibration constants and average noise of all models.}}
                \label{Fig:Averare_model_withData}
 \end{figure}

In Figure~\ref{Fig:Averare_model_withData}, the expected average power, defined in Equation \ref{eq:model}, is shown as a function of time and averaged over frequencies, for the seven sky temperature models. We observe significant variations among different sky models, both in terms of expected power levels and their LST-dependent shapes. The solid line and the gray band will be discussed in the next section.

\section{Calibration method and results}
\label{sec:method}
\hspace{0.5cm} For an accurate interpretation of AERA data, a complete understanding of the entire signal chain (including the combined effects of antenna, amplifiers, filters, and digitizer) is required to minimize measurement uncertainties. Therefore, the antennas and analogue signal chains must be carefully calibrated. To achieve this, we adopted an approach inspired by the calibration technique employed in the LOFAR cosmic ray key science project~\cite{lofar}. This method involves convolving the power emitted by the sky with the gains and noise entering the antenna signal chain, as well as any potential external environmental noise. The frequency-dependent calibration factor, denoted as $C_0(\nu)$, establishes the relationship between the measured signal and the expected output of the antenna, effectively acting as an absolute correction factor. In this way, the model of the power emitted by the sky and propagated through the antenna can be described by
\begin{equation}\label{eq:model_final}
    P_{\mathrm{model}}(t,\nu) = P_{\mathrm{sky}}(t,\nu)G_{\mathrm{ant}}(\nu)G_{\mathrm{AC}}(\nu)C_0^2(\nu) + N_{\mathrm{tot}}^{'}(\nu),
\end{equation}in which $G_{\mathrm{ant}}(\nu)$ and $G_{\mathrm{AC}}(\nu)$ are, respectively, the gains of the  LNA and of the Analogue Chain (AC), where the signal is subjected to a bandpass filter, amplified and, digitized. The free parameters of the model are the calibration constant $C_0(\nu)$ and the total noise $N_{\mathrm{tot}}^{'}(\nu)$, which consists of a sum of the intrinsic electronic thermal noise and the environmental one.

To perform the calibration, we compare the measured signal at the antenna, described by Equation~\ref{eq:power}, with the expected signal given by the sky model $P_{\mathrm{sky}}$. In practice, we use the unfolded data, that is, corrected for detector effects. Therefore, the equation used for the fitting is given by 
\begin{equation}
    P_{\mathrm{model}}(t,\nu)G^{-1}_{\mathrm{AC}}(\nu)G^{-1}_{\mathrm{ant}}(\nu) = P_{\mathrm{sky}}(t,\nu)C_0^2(\nu) + N_{\mathrm{tot}}(\nu),
\end{equation} where the gains of the analog chain and the LNA are applied inversely to the measured signal to remove the effects of the detector. In other words, we compare each bin (in frequency and LST) of the unfolded measured signal, illustrated in the right-hand side of Figure~\ref{Fig:Spectrum_EW_NS} for the north-south channel, with the corresponding bin of the expected power, shown in the right panel of Figure~\ref{Fig:p_model}. Note that $N_{\mathrm{tot}}(\nu)$ now represents the total noise convolved with the inverse gains. For each frequency bin, a linear fit is performed to determine $C_0^2(\nu)$ and $N_{\rm{tot}}(\nu)$. 


\begin{figure}[H]
        \centering
        \includegraphics[scale=0.51]{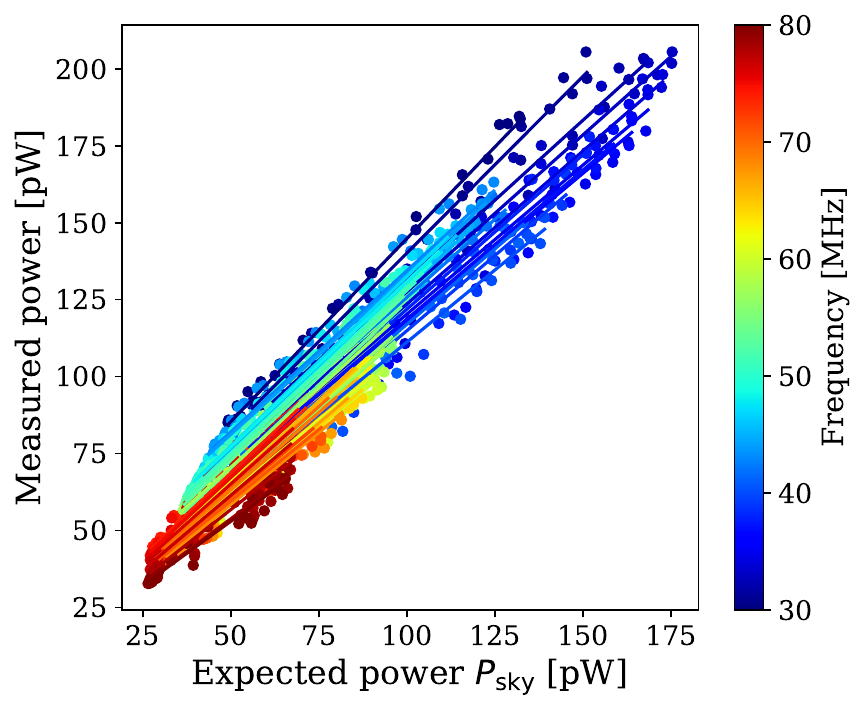}
        \includegraphics[scale=0.51]{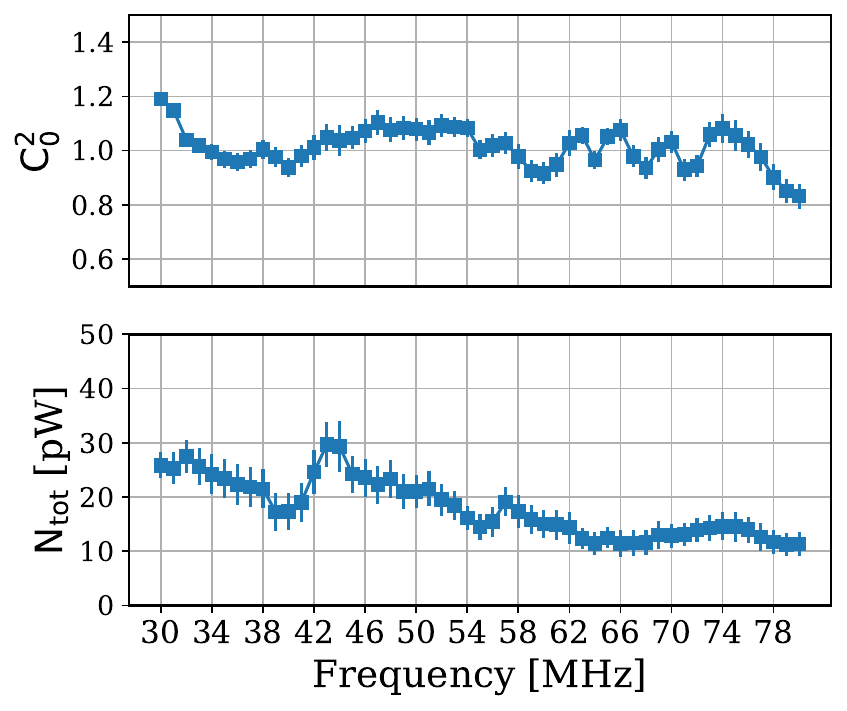}
        \caption{\footnotesize{Left panel: measured power versus the expected power, and the resulting linear fits obtained for each frequency bin. Each fit is based on 24 points corresponding to hourly LST bins. The results correspond to the north-south channel of one of the antennas, using periodic all-triggered traces collected during January 2019. Right panel: calibration constants $C_0^2$ (top panel) and total noise $N_{\rm{tot}}$ (bottom panel) obtained from the fit. 
}}
                \label{Fig:Fit_27}
 \end{figure}
 
\begin{figure}[H]
        \centering
                \includegraphics[scale=0.36]{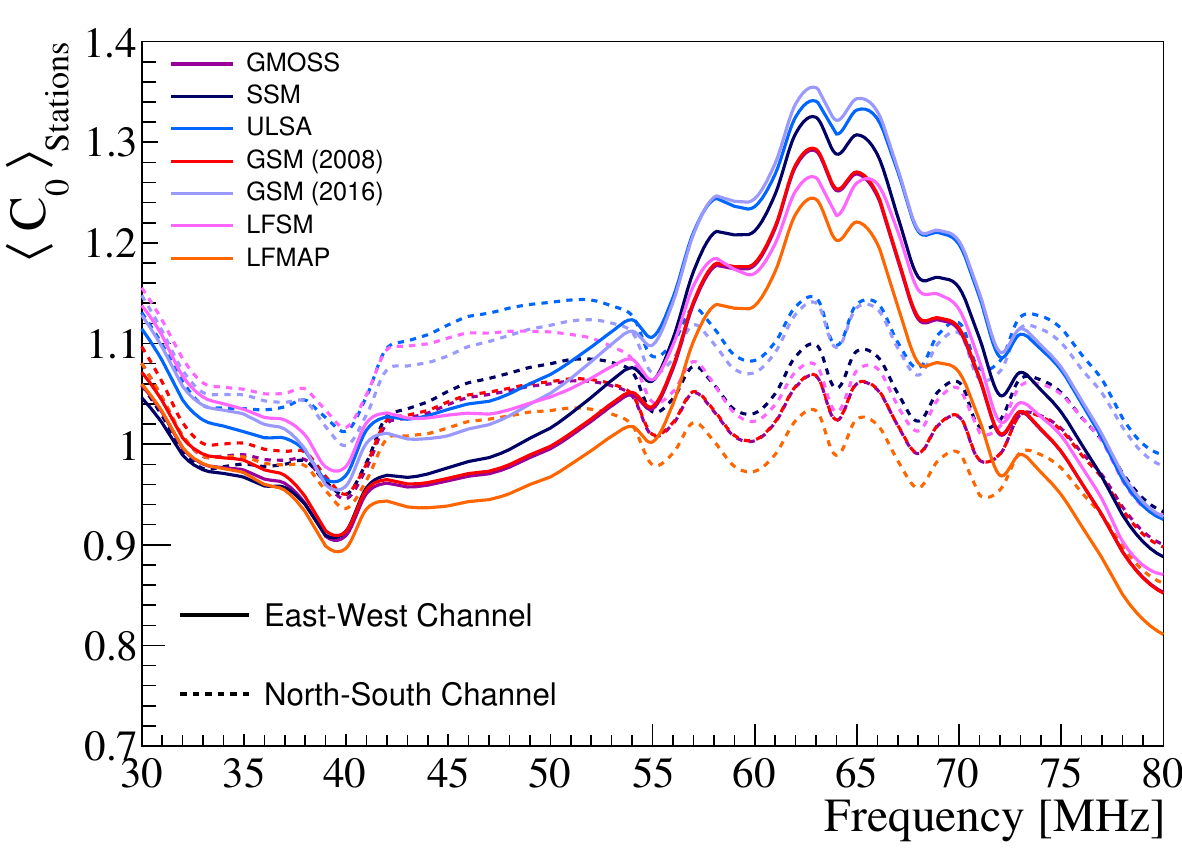} \quad
                \includegraphics[scale=0.36]{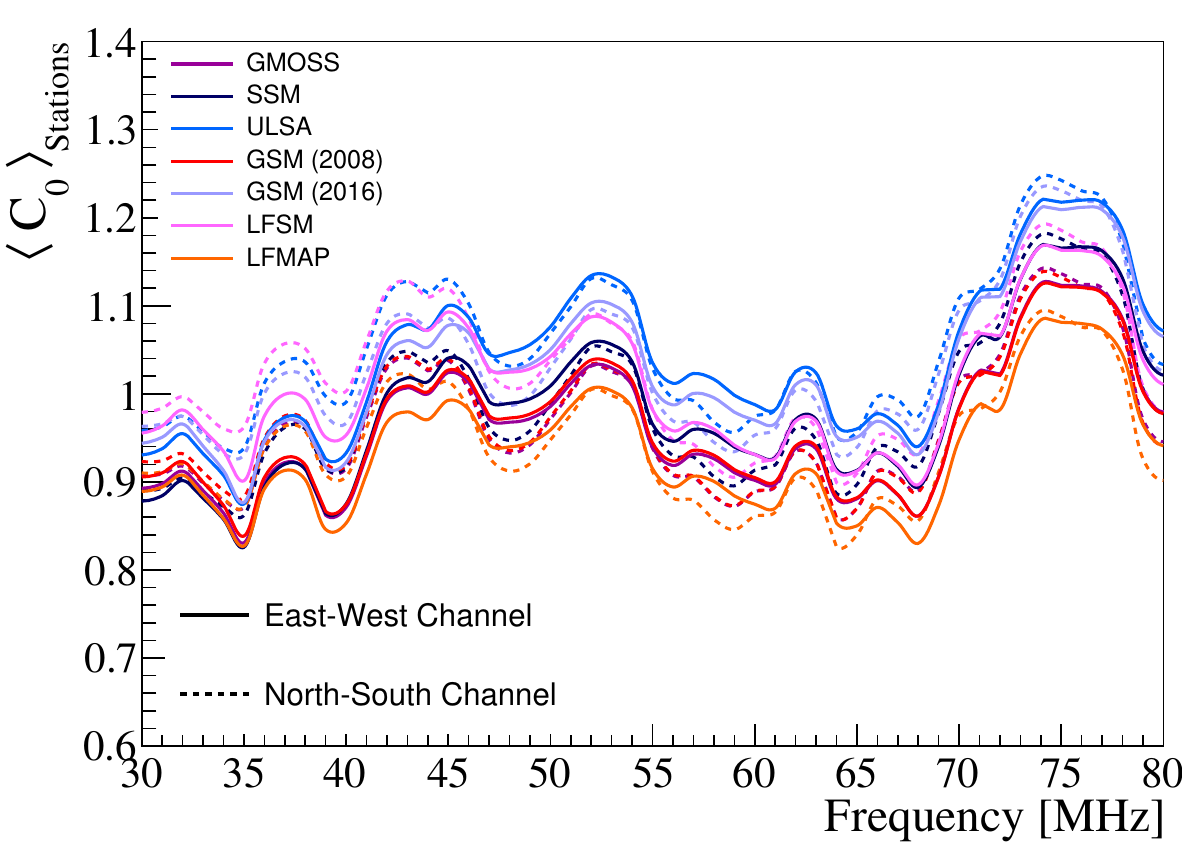} \quad
        \caption{\footnotesize{Average of calibration constants as a function of frequency for all butterfly antennas (left panel) and LPDAs (right panel) obtained for each sky temperature model and sampled monthly over the entire data-taking period described in Section \ref{sec:data_set}. Different models are shown by colored lines, and the two polarization channels are distinguished by dashed (north–south) and solid (east–west) lines. While both channels of the LPDAs exhibit similar behavior, the butterfly antennas show frequency-dependent deviations at higher frequencies due to asymmetries in their directional response.}}
                \label{Fig:results_constants}
 \end{figure}

The calibration for each channel is performed on a monthly basis. Figure~\ref{Fig:Fit_27} shows an example of the fit for each frequency bin for the north–south channel of one of the antennas during January 2019, using the LFmap model. For each frequency bin, the fit is obtained from 24 data points corresponding to the average power measured in hourly LST bins. For this case, the mean $\chi^2$ value obtained from the fits over the different frequency bands is 1.09. Figure~\ref{Fig:results_constants} presents the average calibration constants as a function of frequency, considering all antennas and different sky models. For LPDAs, the north-south and east-west channels exhibit similar behaviors. However, for butterfly antennas, a difference between the north-south and east-west channels is observed around $65$\,MHz, which can be attributed to the directional response asymmetry due to the presence of the electronics box mounted on the antenna pole aligned with the east-west arm. Ideally, these curves would be flat as a function of frequency, and the observed deviations reflect residual uncertainties in the gains and in the modeling of the antenna directional response.

To estimate the impact of the calibration constants on the cosmic ray energy uncertainty, we need to consider a single average calibration constant that takes into account the different radio sky models. We approach the averaging of the calibration constants in two ways. The first approach evaluates the effect for each sky model separately, to estimate the statistical uncertainties within each model and to obtain the systematic spread between models. For this, we first average ${\rm C_0}$ over frequency across the entire frequency band:
\begin{equation}
    \mathbb{C}_0 \equiv \frac{1}{N_{\nu}}\sum_{\nu}  C_0(\nu),
\end{equation}where $N_{\nu}$ is the number of frequency bins. For the computation of $\mathbb{C}_0$, we excluded measurements between $30$\,MHz and $40$\,MHz since we observed significant noise correlated with the solar activity modulation cycle in this frequency range \cite{StudySolarARENA}. These noise contributions consist of radio waves emitted by distant terrestrial sources that are bounced off the atmosphere and arrive at AERA when the Maximum Usable Frequency (MUF), which represents the highest frequency for radio communication between two points on Earth considering ionospheric conditions, is high. In other words,  the atmosphere is not transparent to radio waves for frequencies below the MUF. This typically occurs during periods near the peak of the solar cycle, as the MUF rises in correlation with increased solar activity.  The exclusion of this frequency band is important to avoid distortions in the average constant $\mathbb{C}_0$. Similarly,  the average of the calibration constants $C_0$ reported in Figure~\ref{Fig:results_constants} were computed excluding periods of high solar activity. 
 
The second way to approach the averaging of the calibration constants is to include the variation across the radio sky models in order to obtain a single value that allows evaluation of the evolution over time for each station. For this we follow a similar approach but first average each $C_0(\nu)$ over the models:

\begin{equation}
    \left \langle C_0(\nu) \right \rangle_{\rm{model}} \equiv \frac{1}{N_{\rm{model}}}\sum_{i}  C_{0,i}(\nu),
\end{equation}where  $N_{\rm{model}}=7$ is the number of sky models used in this work and ${\rm{C}_{0,i}(\nu)}$ the calibration constant obtained with each sky model $i$ at frequency $\nu$. 

The solid black line in Figure~\ref{Fig:Averare_model_withData} corresponds to the average power over all frequencies measured by the AERA antennas corrected by the averaged calibration constant $\left \langle C_0(\nu) \right \rangle_{\rm{model}}$. The grey band represents the interval containing the  $95\%$  C.L. of the power measurements from the AERA stations. However, it is important to note that models with expected power outside the $95\%$  C.L. band (the GMOSS model, for example) should not be interpreted as disfavored. This result only means that such models predict powers with significant differences relative to the averaged model. In fact, if the power measured by the AERA antennas were corrected by the calibration constants obtained by using the GMOSS model, the grey band would be centered around the GMOSS model line in Figure~\ref{Fig:Averare_model_withData}. Then we also average over the frequency to obtain the average calibration constant for a station
\begin{equation}
    \left \langle \mathbb{C}_0 \right \rangle_{\rm{model}} = \frac{1}{N_\nu}\sum_{\nu} \left \langle C_0(\nu) \right \rangle_{\rm{model}}.
\end{equation}

\begin{figure}[H]
        \centering
                \includegraphics[scale=0.22]{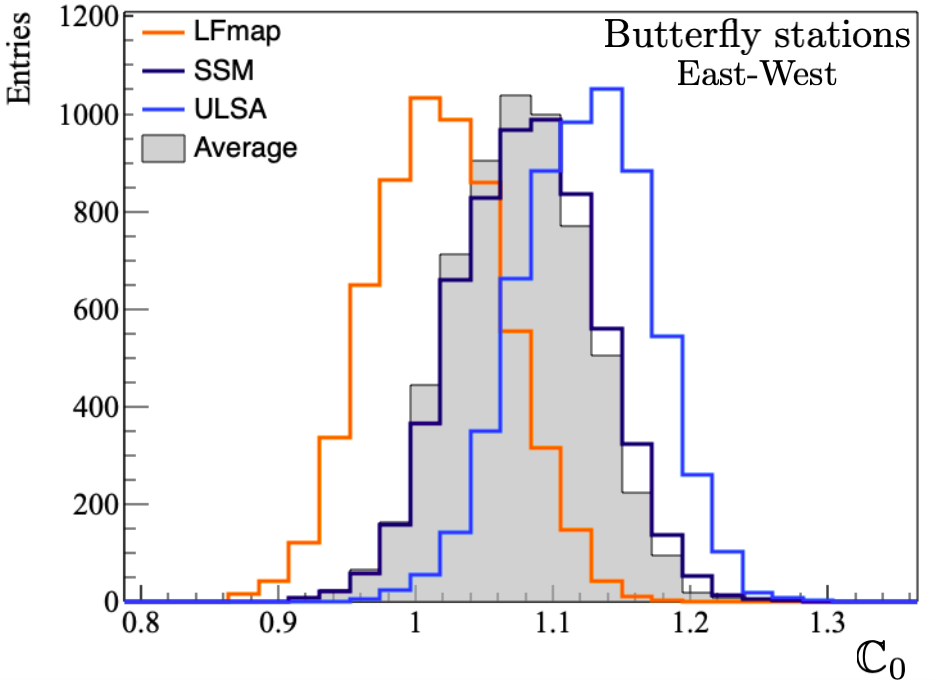} \quad
                \includegraphics[scale=0.22]{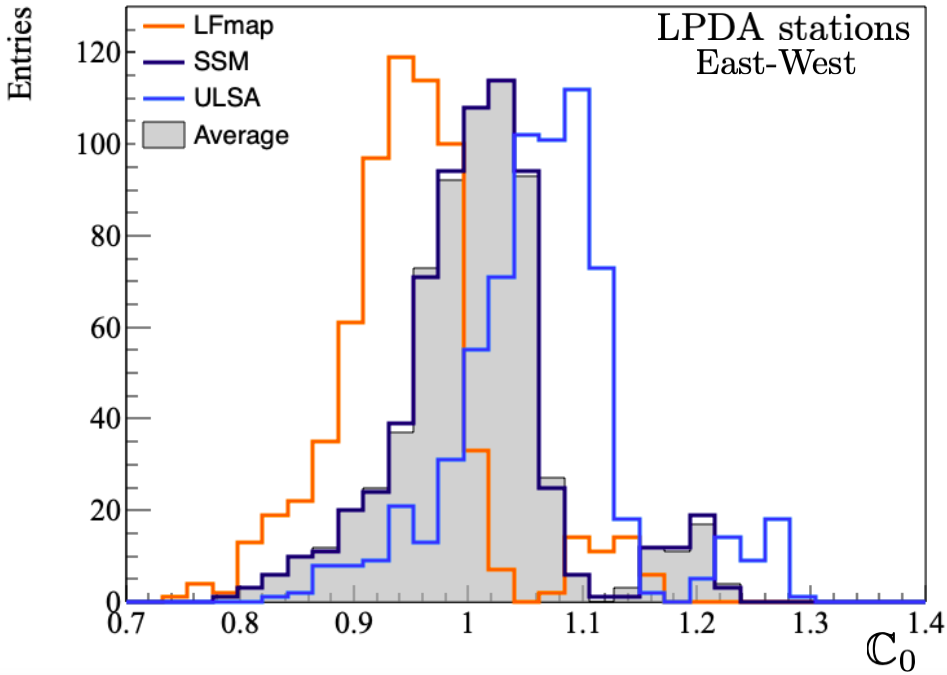} \quad
                \includegraphics[scale=0.22]{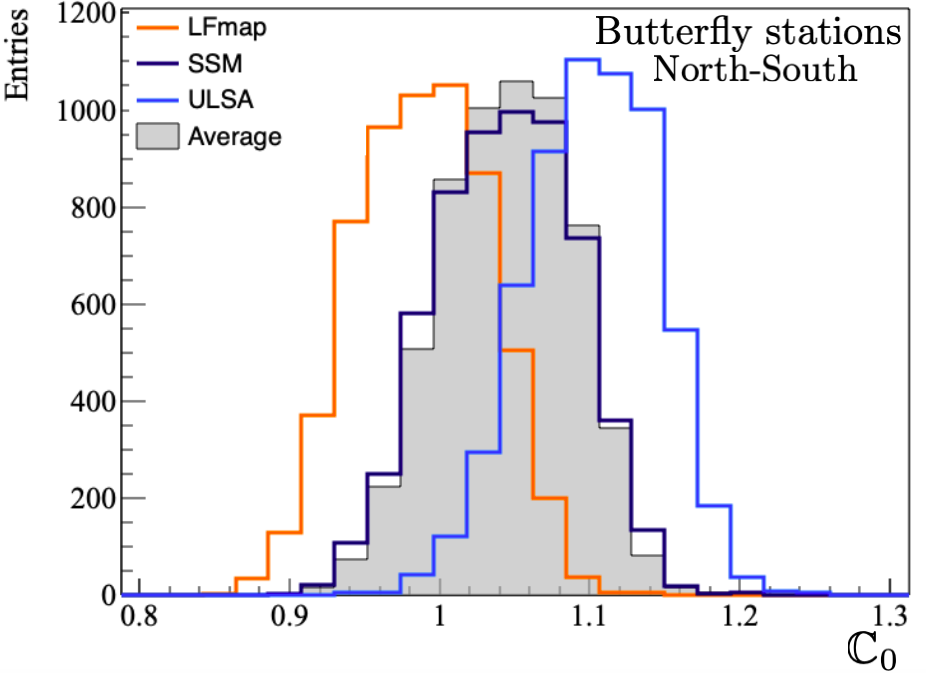} \quad
                \includegraphics[scale=0.22]{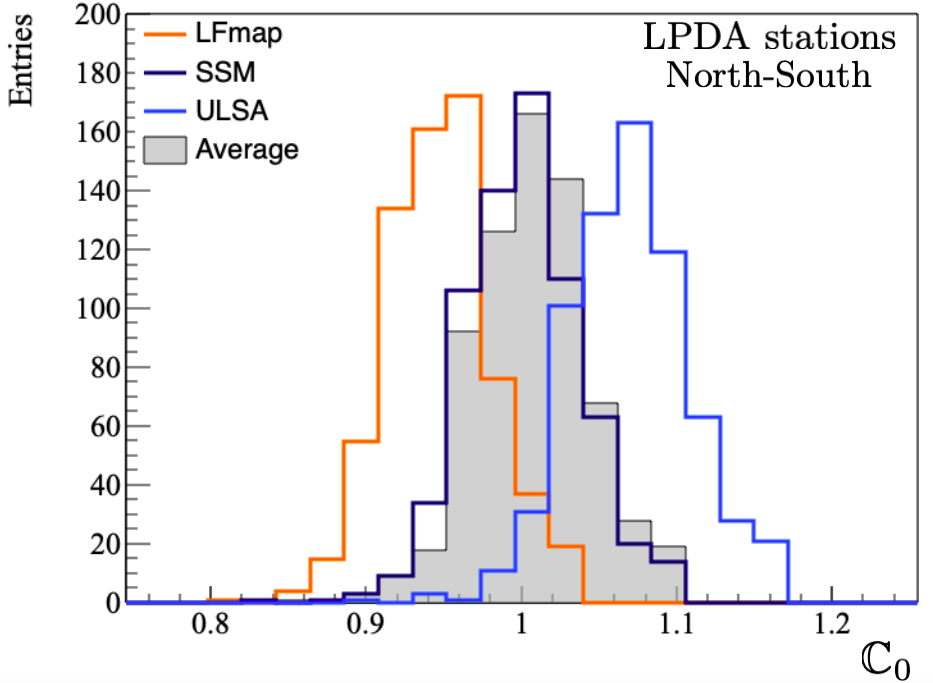} 
        \caption{\footnotesize{Distribution of average calibration constants $\mathbb{C}_0$, obtained from monthly analysis throughout the quality selection period for butterfly antennas (left panels) and LPDAs (right panels). Top: distribution of calibration constants obtained for the east-west channel considering three sky temperature models. Bottom: the same for the north-south channel. The distributions of $\left \langle \mathbb{C}_0 \right \rangle_{\rm{model}}$, denoted by ‘Average’, are also shown in all panels. }}
                \label{Fig:dist_constants}
 \end{figure}

Figure~\ref{Fig:dist_constants} displays the distribution of the averaged calibration constants $\mathbb{C}_0$ obtained for all periods of data taking considering some sky temperature models. The results from the LFmap and ULSA models are presented because they approximately represent the smallest and largest values for the calibration constants, respectively. The SSM model is also included because its distribution lies in an intermediate position among the seven all the considered sky models. In addition, the average over the frequency of the calibration constants previously computed for the different sky models $\left \langle \mathbb{C}_0 \right \rangle_{\rm{model}}$, is also shown.  A small shift of $4\%$ between the north-south and east-west channels in butterfly antennas is observed as a consequence of the imperfect modeling of the electronics box, which has an impact on the directional response of the east-west channel, as seen in Figure~\ref{Fig:results_constants}. The tails observed in the right and left part of the $\mathbb{C}_0$ distribution with respect to the east-west channel of LPDAs arises from measurements related to three specific antennas, where large fluctuation of $\mathbb{C}_0$ values are observed. We reiterate that the number of LPDAs is significantly smaller than the number of butterfly antennas, making their results more susceptible to outliers. The root mean square $ \sigma_{\left \langle \mathbb{C}_0 \right \rangle_{\rm{model}}}$ of the averaged model calibration constant distribution is $\sim 5\%$, primarily driven by gain variations between antennas, event-to-event fluctuations in the background power of each station, and presumably environmental factors. The final estimator, $\widehat{\mathbb{C}}_0$, is summarized in Table~\ref{tab:results_c0} and is obtained by averaging over all stations for each antenna type and polarization channel: 
\begin{equation}
    \widehat{\mathbb{C}}_0 \equiv \left \langle \left \langle \mathbb{C}_0 \right \rangle_{\rm{model}} \right \rangle_{\rm{stations}}.
\end{equation}

As a result of the large number of measurements $N$, equivalent to the number of stations multiplied by the number of months in each dataset, the corresponding  statistical uncertainty $\sigma_{\rm{stat}} = \sigma_{\left \langle \mathbb{C}_0 \right \rangle_{\rm{model}}}/\sqrt{N}$ is negligible for both butterfly antennas ($\lesssim 0.06\%$) and LPDAs ($\lesssim 0.3\%$). Conversely,  the systematic uncertainty $\sigma_{\rm{syst}}$ primarily arises from discrepancies among sky models and is assessed at approximately $6\%$ by computing the mean deviation between the averaged model and LFmap and ULSA models, values that are in good agreement with those estimated in \cite{busken2022}. These two sky models, already discussed above, were selected because they yield the most divergent calibration constant values compared to the average model. Since the cosmic-ray energy is proportional to the square root of the radiation energy~\cite{AERA_energy} (which scales with  $\widehat{\mathbb{C}}_0^2$), the cosmic-ray energy is proportional to  $\widehat{\mathbb{C}}_0$. Therefore, the impact of the systematic uncertainty of the absolute calibration on the radio cosmic-ray energy scale remains at the $6\%$ level.

\begin{table}[h]
\centering
\begin{tabular}{lc}
\hline
\textbf{Station (channel)} & $\widehat{\mathbb{C}}_0$ $\boldsymbol{\pm 
 \ \sigma_{\text{syst}}}$ \\
\hline
butterfly (east-west) & $1.08 \pm 0.05$ \\
butterfly (north-south) & $1.04 \pm 0.06$ \\
LPDA (east-west) & $1.01 \pm 0.06$ \\
LPDA (north-south) & $1.01 \pm 0.06$ \\
\hline
\end{tabular}
\caption{Calibration constant results obtained for both channels of butterfly and LPDA antennas.}
\label{tab:results_c0}
\end{table}

\section{Study of the calibration constant as a function of time}
\label{sec:c0_time}

\begin{figure}[H]
        \centering
                \includegraphics[scale=0.39]{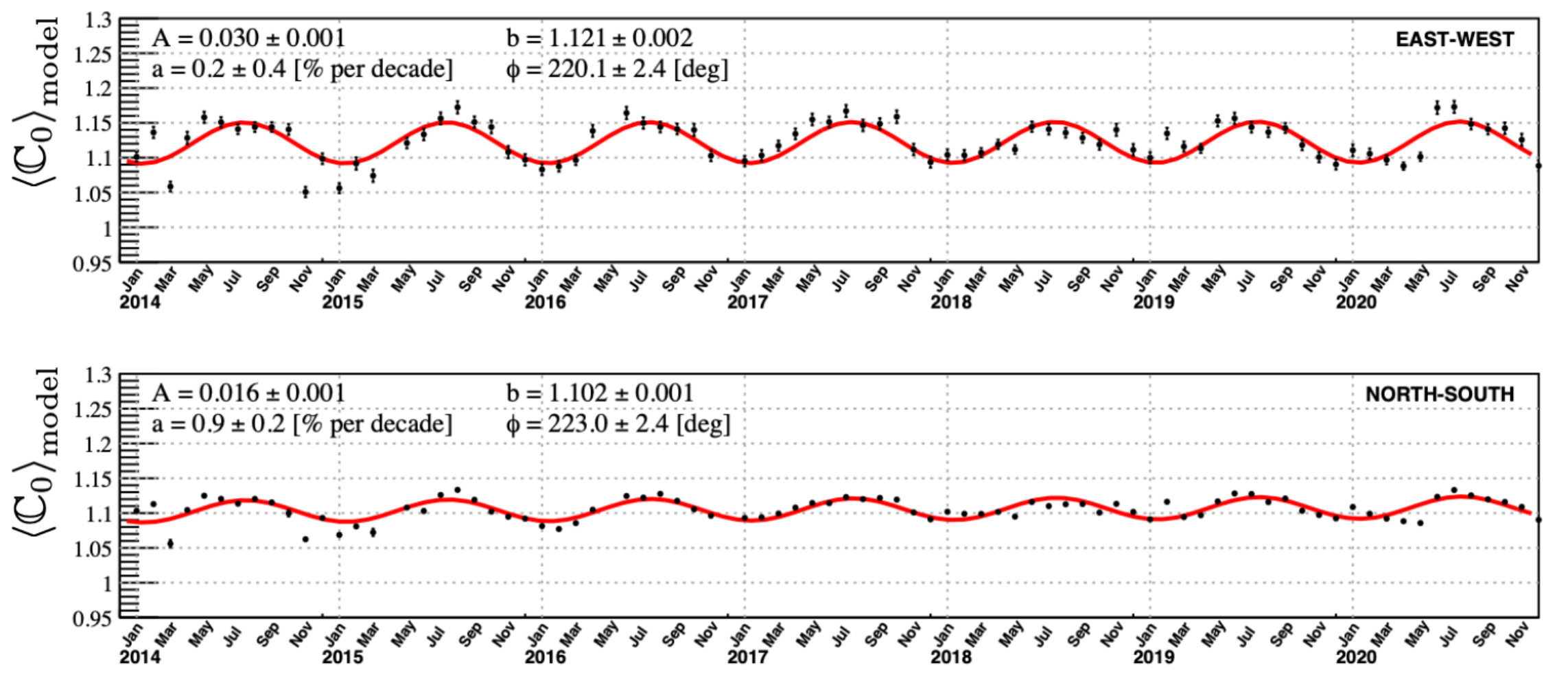}
        \caption{\footnotesize{Calibration constants obtained for both channels of antenna Id:33 from 2014 to 2020. The corresponding
cosine+linear fit is represented by the red curve.}}
                \label{Fig:C0_time}
 \end{figure}

\hspace{0.5cm} In this section, we study the time evolution of the calibration constants over the entire data-taking period. For this, we consider, for each antenna, the temporal evolution of the  $\left \langle \mathbb{C}_0 \right \rangle_{\rm{model}}$ obtained values. In Figure~\ref{Fig:C0_time} we show, as an example, the behavior of the $\left \langle \mathbb{C}_0 \right \rangle_{\rm{model}}$ values for both channels of a specific butterfly antenna from 2014 until the end of 2020. Note that even after correcting for the temperature-dependent gain variations of amplifiers in the signal chain, a residual seasonal modulation is observed. This variation can be primarily attributed to the effect of changing noise conditions over time. The method used to extract the $C_0$ and noise term assumes the noise is constant over time, however, this assumption is subtly broken when correcting for the temperature-dependent gain. In this correction, the temperature dependence of the LNA is applied to the total measured signal, but this affects not only the signal and noise before the LNA but also the system noise. As a result, the temperature correction removes to a large extent the temperature dependence of the measured signal, but a (smaller) noise dependence is introduced. This results in a small mixing between changes in the noise and the Galactic signal, leading to slightly different slopes in the fit of measured versus expected power. The seasonal modulation then arises because the galaxy drifts over the year with sidereal time, while the noise changes on the day-night cycle. The effect has been verified by comparing $C_0$ values for day and night separately. This comparison shows a small difference in the slope on the order of the observed seasonal modulation over the year. Note that this effect averages out over a year hence it does not affect the final absolute calibration of the detector.  Besides, as discussed earlier, the approximation of the LNA temperature by the ambient temperature recorded at the CRS (see section \ref{sec:data_set}) could also be one potential factor contributing to the observed modulation. Additionally, other factors such as wet ground conditions or mismatches attributable to the sky models cannot be disregarded, but these effects appear to be subdominant to the larger modulation introduced by the temperature compensation.

Accounting for this modulation,  the time evolution of the calibration constant is parameterized as 
\begin{equation}
    \left \langle \mathbb{C}_0(t') \right \rangle_{\rm{model}} = A\cos{(\frac{\pi}{6}t' + \phi)} + at' + b,
\end{equation}where the parameters $A$ and $\phi$ represent the magnitude and phase of the observed seasonal modulation, respectively. $b$ denotes the baseline value of the calibration constant. The variable $t'$ denotes the time in months since the start of data taking. The slope parameter $a$ is particularly significant, as it represents the monthly ageing rate of the AERA detector station. An illustration of the fit of the calibration constants over time is depicted by the red curve in Figure~\ref{Fig:C0_time}.

\begin{figure}[H]
        \centering
                \includegraphics[scale=0.36]{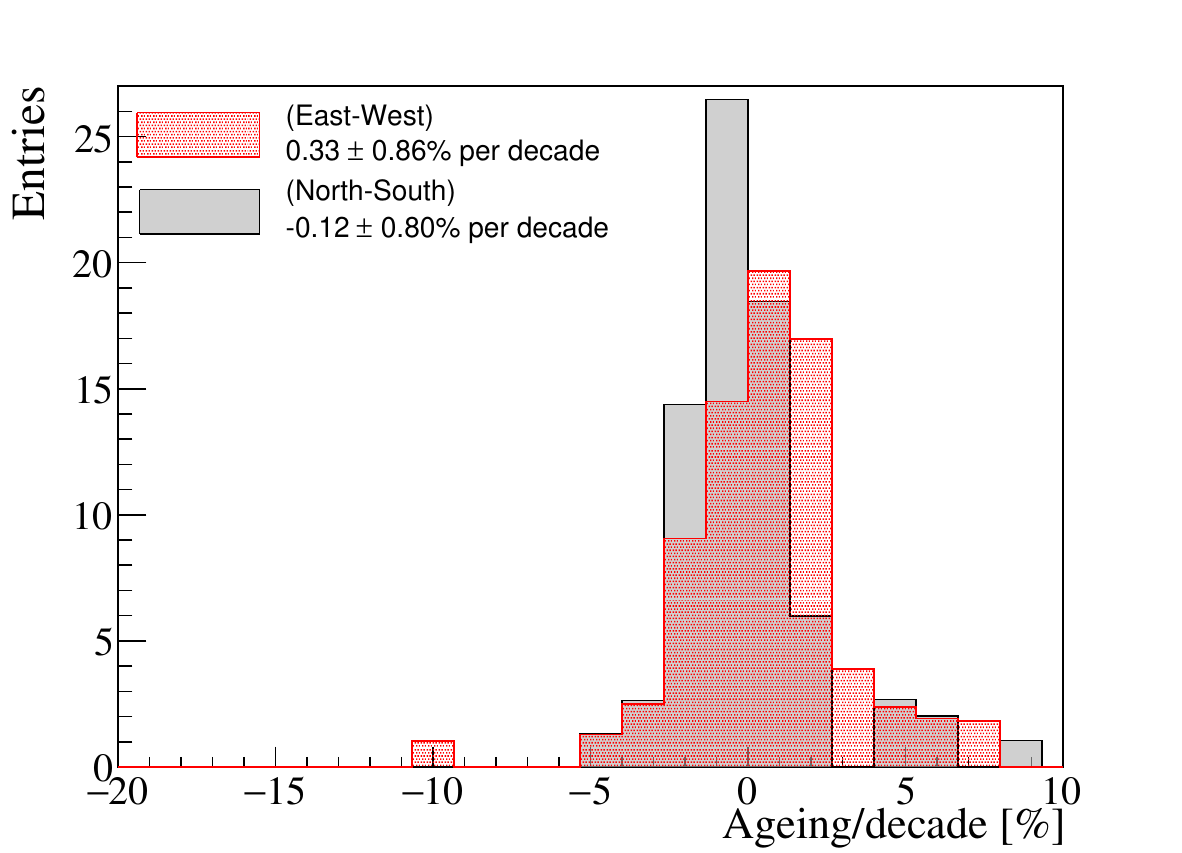} \quad 
                \includegraphics[scale=0.36]{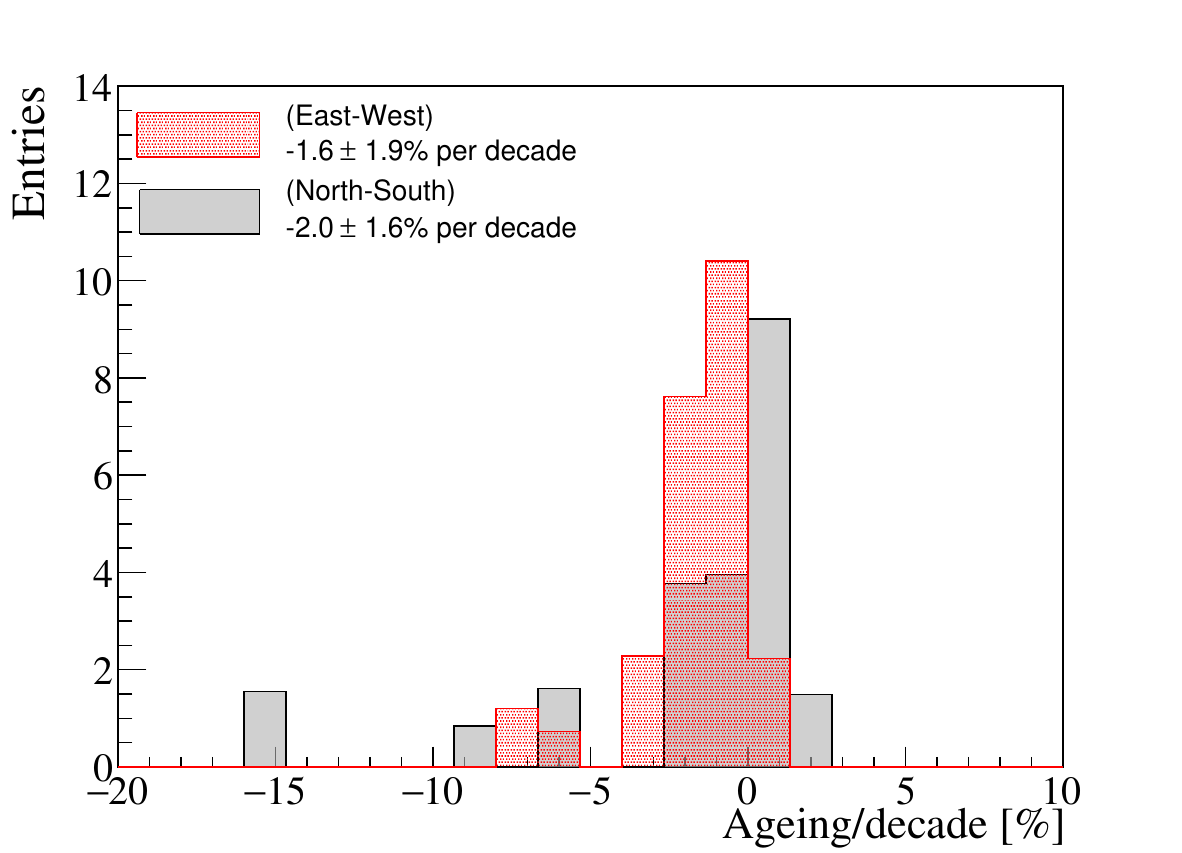} \quad 
        \caption{\footnotesize{Distribution of the coefficients $a$ (interpreted as the ageing factor) in percentage per decade, obtained from each antenna and channel and averaged over frequency bins. Left: butterfly stations. Right: LPDAs.}}
                \label{Fig:aging_results}
 \end{figure}

The fitting is done for all channels considered in this study. In Figure~\ref{Fig:aging_results} we present the distribution of ageing factors $a$ per decade for both channels, considering the butterfly antennas and LPDAs in the left and right panels, respectively. Regarding the uncertainty estimation of the ageing factor, it is important to note that the fit does not describe the seasonal modulation perfectly. There are upward/downward fluctuations of the values of $\left \langle \mathbb{C}_0 \right \rangle_{\rm{model}}$ observed for all antennas at the same time in some specific periods that potentially impact the resulting fitted ageing coefficient. The impact is particularly pronounced when significant fluctuations coincide with the beginning or end of the data collection period. Therefore, it is crucial to not confuse genuine ageing effects with transient fluctuations that happen to occur during later or earlier years. To address this challenge, we conducted mock simulations of the calibration constants over time. In these simulations, for each antenna and month, we generated random Gaussian-distributed values of calibration constants with mean and standard deviation equal to the measured $\left \langle \mathbb{C}_0 \right \rangle_{\rm{model}}$ value and its corresponding uncertainty. Subsequently, we systematically shuffle the years for all antennas and perform fits on the simulated calibration constants over time. The average of the mock ageing coefficients should converge to zero, and the standard deviation of the mock ageing distribution serves as our final estimate for the uncertainty in the ageing parameter.

This approach accounts for significant fluctuations that occurred during specific periods, which may be unrelated to detector ageing. The final results for the ageing coefficient per decade, separated per antenna and channel, are summarized in Table~\ref{tab:results_aging}. Notice that the results from the LPDAs are more prone to large fluctuations because the statistics are lower (only 14 antennas) and a shorter data collection period (4 years) was used.  Combining all antenna types and channels, we obtain an ageing factor of $a = (-0.27 \pm 0.53) \%$ per decade, indicating that no significant or relevant ageing effect is observed in the AERA detector stations.

\begin{table}[h]
\centering
\begin{tabular}{lcc}
\hline
\textbf{Station (channel)} & \textbf{Ageing per decade} [\%] \\
\hline
butterfly (east-west) & $0.33 \pm 0.86$ \\
butterfly (north-south) & $-0.12 \pm 0.80$ \\
LPDA (east-west) & $-1.6 \pm 1.9$\\
LPDA (north-south) & $-2.0 \pm 1.6$\\
\hline
\end{tabular}
\caption{Ageing factor for each channel per decade.}
\label{tab:results_aging}
\end{table}

  \section{Conclusions} 
\label{sec:conclusions}
\hspace{0.5cm}  In this study, we performed an absolute frequency-dependent Galactic calibration of AERA stations, encompassing both LPDAs and butterfly antennas. By using seven different radio sky models, we obtained the average calibration constant $\widehat{\mathbb{C}}_0$ for each channel and the corresponding statistical and systematic uncertainties. The estimated calibration constants derived are consistent with $\widehat{\mathbb{C}}_0 = 1$ within uncertainties, indicating very good agreement with the original calibration process. This process involved laboratory measurements of the analogue chain, as well as simulations and analysis to ascertain the directional response of the antennas. Furthermore, we investigated the temporal behavior of the calibration constants over a period of seven years. The ageing coefficients deduced from fitting the evolution over time of the  calibration constants were found to be negligible and consistent with zero within the uncertainties. This indicates the absence of significant ageing effects in the AERA detector stations and demonstrates the possibility for a radio detector to effectively monitor ageing effects in other detectors operating over extended periods. These results are particularly valuable in the context of the Pierre Auger Observatory upgrade~\cite{AugerPrime} and the Auger Radio Detector~\cite{RD-paper}, highlighting the importance of the radio detection technique for determining an absolute energy scale for cosmic rays.




\clearpage
\section*{The Pierre Auger Collaboration}
\small

\begin{sloppypar}\noindent
\input{latex_authorlist_authors}
\end{sloppypar}

\begin{center}
\rule{0.1\columnwidth}{0.5pt}
\raisebox{-0.4ex}{\scriptsize$\bullet$}
\rule{0.1\columnwidth}{0.5pt}
\end{center}

\vspace{-1ex}
\footnotesize
\input{latex_authorlist_institutions}

\vspace{-1ex}
\footnotesize
\input{acknowledgments}

\end{document}

%% file: latex_authorlist_authors.tex
A.~Abdul Halim$^{13}$,
P.~Abreu$^{67}$,
M.~Aglietta$^{50,49}$,
I.~Allekotte$^{1}$,
K.~Almeida Cheminant$^{75,74}$,
R.~Aloisio$^{42,43}$,
J.~Alvarez-Mu\~niz$^{73}$,
A.~Ambrosone$^{42}$,
J.~Ammerman Yebra$^{73}$,
L.~Anchordoqui$^{79}$
% ...

%% file: latex_authorlist_institutions.tex
\begin{description}[labelsep=0.2em,align=right,labelwidth=0.7em,labelindent=0em,leftmargin=2em,noitemsep,before={\renewcommand\makelabel[1]{##1 }}]
\item[$^{1}$] Centro At\'omico Bariloche and Instituto Balseiro (CNEA-UNCuyo-CONICET), San Carlos de Bariloche, Argentina
\item[$^{2}$] Departamento de F\'\i{}sica and Departamento de Ciencias de la Atm\'osfera y los Oc\'eanos, FCEyN, Universidad de Buenos Aires and CONICET, Buenos Aires, Argentina
\item[$^{3}$] IFLP, Universidad Nacional de La Plata and CONICET, La Plata, Argentina
\item[$^{4}$] Instituto de Astronom\'\i{}a y F\'\i{}sica del Espacio (IAFE, CONICET-UBA), Buenos Aires, Argentina
\item[$^{5}$] Instituto de F\'\i{}sica de Rosario (IFIR) -- CONICET/U.N.R.\ and Facultad de Ciencias Bioqu\'\i{}micas y Farmac\'euticas U.N.R., Rosario, Argentina
\item[$^{6}$] Instituto de Tecnolog\'\i{}as en Detecci\'on y Astropart\'\i{}culas (CNEA, CONICET, UNSAM), and Universidad Tecnol\'ogica Nacional -- Facultad Regional Mendoza (CONICET/CNEA), Mendoza, Argentina
\item[$^{7}$] Instituto de Tecnolog\'\i{}as en Detecci\'on y Astropart\'\i{}culas (CNEA, CONICET, UNSAM), Buenos Aires, Argentina
\item[$^{8}$] International Center of Advanced Studies and Instituto de Ciencias F\'\i{}sicas, ECyT-UNSAM and CONICET, Campus Miguelete -- San Mart\'\i{}n, Buenos Aires, Argentina
\item[$^{9}$] Laboratorio Atm\'osfera -- Departamento de Investigaciones en L\'aseres y sus Aplicaciones -- UNIDEF (CITEDEF-CONICET), Argentina
\item[$^{10}$] Observatorio Pierre Auger, Malarg\"ue, Argentina
\item[$^{11}$] Observatorio Pierre Auger and Comisi\'on Nacional de Energ\'\i{}a At\'omica, Malarg\"ue, Argentina
\item[$^{12}$] Universidad Tecnol\'ogica Nacional -- Facultad Regional Buenos Aires, Buenos Aires, Argentina
\item[$^{13}$] University of Adelaide, Adelaide, S.A., Australia
\item[$^{14}$] Universit\'e Libre de Bruxelles (ULB), Brussels, Belgium
\item[$^{15}$] Vrije Universiteit Brussels, Brussels, Belgium
\item[$^{16}$] Centro Brasileiro de Pesquisas Fisicas, Rio de Janeiro, RJ, Brazil
\item[$^{17}$] Centro Federal de Educa\c{c}\~ao Tecnol\'ogica Celso Suckow da Fonseca, Petropolis, Brazil
\item[$^{18}$] Universidade de S\~ao Paulo, Escola de Engenharia de Lorena, Lorena, SP, Brazil
\item[$^{19}$] Universidade de S\~ao Paulo, Instituto de F\'\i{}sica de S\~ao Carlos, S\~ao Carlos, SP, Brazil
\item[$^{20}$] Universidade de S\~ao Paulo, Instituto de F\'\i{}sica, S\~ao Paulo, SP, Brazil
\item[$^{21}$] Universidade Estadual de Campinas (UNICAMP), IFGW, Campinas, SP, Brazil
\item[$^{22}$] Universidade Estadual de Feira de Santana, Feira de Santana, Brazil
\item[$^{23}$] Universidade Federal do ABC, Santo Andr\'e, SP, Brazil
\item[$^{24}$] Universidade Federal do Paran\'a, Setor Palotina, Palotina, Brazil
\item[$^{25}$] Universidade Federal do Rio de Janeiro, Instituto de F\'\i{}sica, Rio de Janeiro, RJ, Brazil
\item[$^{26}$] Universidad de Medell\'\i{}n, Medell\'\i{}n, Colombia
\item[$^{27}$] Universidad Industrial de Santander, Bucaramanga, Colombia
\item[$^{28}$] Charles University, Faculty of Mathematics and Physics, Institute of Particle and Nuclear Physics, Prague, Czech Republic
\item[$^{29}$] Institute of Physics of the Czech Academy of Sciences, Prague, Czech Republic
\item[$^{30}$] Palacky University, Olomouc, Czech Republic
\item[$^{31}$] CNRS/IN2P3, IJCLab, Universit\'e Paris-Saclay, Orsay, France
\item[$^{32}$] Laboratoire de Physique Nucl\'eaire et de Hautes Energies (LPNHE), Sorbonne Universit\'e, Universit\'e de Paris, CNRS-IN2P3, Paris, France
\item[$^{33}$] Universit\'e Paris-Saclay, CNRS/IN2P3, IJCLab, Orsay, France
\item[$^{34}$] Bergische Universit\"at Wuppertal, Department of Physics, Wuppertal, Germany
\item[$^{35}$] Karlsruhe Institute of Technology (KIT), Institute for Experimental Particle Physics, Karlsruhe, Germany
\item[$^{36}$] Karlsruhe Institute of Technology (KIT), Institut f\"ur Prozessdatenverarbeitung und Elektronik, Karlsruhe, Germany
\item[$^{37}$] Karlsruhe Institute of Technology (KIT), Institute for Astroparticle Physics, Karlsruhe, Germany
\item[$^{38}$] RWTH Aachen University, III.\ Physikalisches Institut A, Aachen, Germany
\item[$^{39}$] TU Dortmund University, Department of Physics, Dortmund, Germany
\item[$^{40}$] Universit\"at Hamburg, II.\ Institut f\"ur Theoretische Physik, Hamburg, Germany
\item[$^{41}$] Universit\"at Siegen, Department Physik -- Experimentelle Teilchenphysik, Siegen, Germany
\item[$^{42}$] Gran Sasso Science Institute, L'Aquila, Italy
\item[$^{43}$] INFN Laboratori Nazionali del Gran Sasso, Assergi (L'Aquila), Italy
\item[$^{44}$] INFN, Sezione di Catania, Catania, Italy
\item[$^{45}$] INFN, Sezione di Lecce, Lecce, Italy
\item[$^{46}$] INFN, Sezione di Milano, Milano, Italy
\item[$^{47}$] INFN, Sezione di Napoli, Napoli, Italy
\item[$^{48}$] INFN, Sezione di Roma ``Tor Vergata'', Roma, Italy
\item[$^{49}$] INFN, Sezione di Torino, Torino, Italy
\item[$^{50}$] Osservatorio Astrofisico di Torino (INAF), Torino, Italy
\item[$^{51}$] Politecnico di Milano, Dipartimento di Scienze e Tecnologie Aerospaziali , Milano, Italy
\item[$^{52}$] Universit\`a del Salento, Dipartimento di Matematica e Fisica ``E.\ De Giorgi'', Lecce, Italy
\item[$^{53}$] Universit\`a dell'Aquila, Dipartimento di Scienze Fisiche e Chimiche, L'Aquila, Italy
\item[$^{54}$] Universit\`a di Catania, Dipartimento di Fisica e Astronomia ``Ettore Majorana``, Catania, Italy
\item[$^{55}$] Universit\`a di Milano, Dipartimento di Fisica, Milano, Italy
\item[$^{56}$] Universit\`a di Napoli ``Federico II'', Dipartimento di Fisica ``Ettore Pancini'', Napoli, Italy
\item[$^{57}$] Universit\`a di Palermo, Dipartimento di Fisica e Chimica ''E.\ Segr\`e'', Palermo, Italy
\item[$^{58}$] Universit\`a di Roma ``Tor Vergata'', Dipartimento di Fisica, Roma, Italy
\item[$^{59}$] Universit\`a Torino, Dipartimento di Fisica, Torino, Italy
\item[$^{60}$] Benem\'erita Universidad Aut\'onoma de Puebla, Puebla, M\'exico
\item[$^{61}$] Unidad Profesional Interdisciplinaria en Ingenier\'\i{}a y Tecnolog\'\i{}as Avanzadas del Instituto Polit\'ecnico Nacional (UPIITA-IPN), M\'exico, D.F., M\'exico
\item[$^{62}$] Universidad Aut\'onoma de Chiapas, Tuxtla Guti\'errez, Chiapas, M\'exico
\item[$^{63}$] Universidad Michoacana de San Nicol\'as de Hidalgo, Morelia, Michoac\'an, M\'exico
\item[$^{64}$] Universidad Nacional Aut\'onoma de M\'exico, M\'exico, D.F., M\'exico
\item[$^{65}$] Institute of Nuclear Physics PAN, Krakow, Poland
\item[$^{66}$] University of \L{}\'od\'z, Faculty of High-Energy Astrophysics,\L{}\'od\'z, Poland
\item[$^{67}$] Laborat\'orio de Instrumenta\c{c}\~ao e F\'\i{}sica Experimental de Part\'\i{}culas -- LIP and Instituto Superior T\'ecnico -- IST, Universidade de Lisboa -- UL, Lisboa, Portugal
\item[$^{68}$] ``Horia Hulubei'' National Institute for Physics and Nuclear Engineering, Bucharest-Magurele, Romania
\item[$^{69}$] Institute of Space Science, Bucharest-Magurele, Romania
\item[$^{70}$] Center for Astrophysics and Cosmology (CAC), University of Nova Gorica, Nova Gorica, Slovenia
\item[$^{71}$] Experimental Particle Physics Department, J.\ Stefan Institute, Ljubljana, Slovenia
\item[$^{72}$] Universidad de Granada and C.A.F.P.E., Granada, Spain
\item[$^{73}$] Instituto Galego de F\'\i{}sica de Altas Enerx\'\i{}as (IGFAE), Universidade de Santiago de Compostela, Santiago de Compostela, Spain
\item[$^{74}$] IMAPP, Radboud University Nijmegen, Nijmegen, The Netherlands
\item[$^{75}$] Nationaal Instituut voor Kernfysica en Hoge Energie Fysica (NIKHEF), Science Park, Amsterdam, The Netherlands
\item[$^{76}$] Stichting Astronomisch Onderzoek in Nederland (ASTRON), Dwingeloo, The Netherlands
\item[$^{77}$] Case Western Reserve University, Cleveland, OH, USA
\item[$^{78}$] Colorado School of Mines, Golden, CO, USA
\item[$^{79}$] Department of Physics and Astronomy, Lehman College, City University of New York, Bronx, NY, USA
\item[$^{80}$] Michigan Technological University, Houghton, MI, USA
\item[$^{81}$] New York University, New York, NY, USA
\item[$^{82}$] University of Chicago, Enrico Fermi Institute, Chicago, IL, USA
\item[$^{83}$] University of Delaware, Department of Physics and Astronomy, Bartol Research Institute, Newark, DE, USA
\item[] -----
\item[$^{a}$] Max-Planck-Institut f\"ur Radioastronomie, Bonn, Germany
\item[$^{b}$] also at Kapteyn Institute, University of Groningen, Groningen, The Netherlands
\item[$^{c}$] School of Physics and Astronomy, University of Leeds, Leeds, United Kingdom
\item[$^{d}$] Fermi National Accelerator Laboratory, Fermilab, Batavia, IL, USA
\item[$^{e}$] Pennsylvania State University, University Park, PA, USA
\item[$^{f}$] Colorado State University, Fort Collins, CO, USA
\item[$^{g}$] Louisiana State University, Baton Rouge, LA, USA
\item[$^{h}$] Universit\'e Grenoble Alpes, CNRS, Grenoble Institute of Engineering, LPSC-IN2P3, Grenoble, France
\item[$^{i}$] now at Graduate School of Science, Osaka Metropolitan University, Osaka, Japan
\item[$^{j}$] Institut universitaire de France (IUF), France
\item[$^{k}$] now at Technische Universit\"at Dortmund and Ruhr-Universit\"at Bochum, Dortmund and Bochum, Germany
\end{description}

%% file: acknowledgments.tex
\section*{Acknowledgments}

\begin{sloppypar}
The successful installation, commissioning, and operation of the Pierre
Auger Observatory would not have been possible without the strong
commitment and effort from the technical and administrative staff in
Malarg\"ue. We are very grateful to the following agencies and
organizations for financial support:
\end{sloppypar}

\begin{sloppypar}
Argentina -- Comisi\'on Nacional de Energ\'\i{}a At\'omica; Agencia Nacional de
Promoci\'on Cient\'\i{}fica y Tecnol\'ogica (ANPCyT); Consejo Nacional de
Investigaciones Cient\'\i{}ficas y T\'ecnicas (CONICET); Gobierno de la
Provincia de Mendoza; Municipalidad de Malarg\"ue; NDM Holdings and Valle
Las Le\~nas; in gratitude for their continuing cooperation over land
access; Australia -- the Australian Research Council; Belgium -- Fonds
de la Recherche Scientifique (FNRS); Research Foundation Flanders (FWO),
Marie Curie Action of the European Union Grant No.~101107047; Brazil --
Minist\'erio da Ci\^encia, Tecnologia e Inova\c{c}\~ao (MCTI); Czech Republic --
GACR 24-13049S, CAS LQ100102401, MEYS LM2023032,
CZ.02.1.01/0.0/0.0/16{\textunderscore}013/0001402, CZ.02.1.01/0.0/0.0/18{\textunderscore}046/0016010
and CZ.02.1.01/0.0/0.0/17{\textunderscore}049/0008422 and
CZ.02.01.01/00/22{\textunderscore}008/0004632; France -- Centre de Calcul IN2P3/CNRS;
Centre National de la Recherche Scientifique (CNRS); Institut National
de Physique Nucl\'eaire et de Physique des Particules (IN2P3/CNRS);
Germany -- Bundesministerium f\"ur Forschung, Technologie und Raumfahrt
(BMFTR); Deutsche Forschungsgemeinschaft (DFG); Ministerium f\"ur Finanzen
Baden-W\"urttemberg; Helmholtz Alliance for Astroparticle Physics (HAP);
Hermann von Helmholtz-Gemeinschaft Deutscher Forschungszentren e.V.;
Ministerium f\"ur Kultur und Wissenschaft des Landes Nordrhein-Westfalen;
Ministerium f\"ur Wissenschaft, Forschung und Kunst des Landes
Baden-W\"urttemberg; Italy -- Istituto Nazionale di Fisica Nucleare
(INFN); Istituto Nazionale di Astrofisica (INAF); Ministero
dell'Universit\`a e della Ricerca (MUR); CETEMPS Center of Excellence;
Ministero degli Affari Esteri (MAE), ICSC Centro Nazionale di Ricerca in
High Performance Computing, Big Data and Quantum Computing, funded by
European Union NextGenerationEU, reference code CN{\textunderscore}00000013; M\'exico --
Consejo Nacional de Ciencia y Tecnolog\'\i{}a (CONACYT-SECHTI)
No.~CB-A1-S-46703, Universidad Nacional Aut\'onoma de M\'exico (UNAM)
PAPIIT-IN114924; Benem\'erita Universidad Aut\'onoma de Puebla (BUAP), VIEP
and Laboratorio Nacional de Superc\'omputo del Sureste de M\'exico (LNS);
and Benem\'erita Universidad Aut\'onoma de Chiapas (UNACH); The Netherlands
-- Ministry of Education, Culture and Science; Netherlands Organisation
for Scientific Research (NWO); Dutch national e-infrastructure with the
support of SURF Cooperative; Poland -- Ministry of Science and Higher
Education, grant No.~2022/WK/12; National Science Centre, grants
No.~2020/39/B/ST9/01398, and 2022/45/B/ST9/02163; Portugal -- Portuguese
national funds and FEDER funds within Programa Operacional Factores de
Competitividade through Funda\c{c}\~ao para a Ci\^encia e a Tecnologia
(COMPETE); Romania -- Ministry of Education and Research, contract
no.~30N/2023 under Romanian National Core Program LAPLAS VII, and grant
no.~PN 23 21 01 02; Slovenia -- Slovenian Research and Innovation
Agency, grants P1-0031, I0-0033; Spain -- Ministerio de Ciencia,
Innovaci\'on y Universidades/Agencia Estatal de Investigaci\'on MICIU/AEI
/10.13039/501100011033 (PID2022-140510NB-I00, PCI2023-145952-2,
CNS2024-154676, and Mar\'\i{}a de Maeztu CEX2023-001318-M), Xunta de Galicia
(CIGUS Network of Research Centers, Consolidaci\'on ED431C-2025/11 and
ED431F-2022/15) and European Union ERDF; USA -- Department of Energy,
Contracts No.~DE-AC02-07CH11359, No.~DE-FR02-04ER41300,
No.~DE-FG02-99ER41107 and No.~DE-SC0011689; National Science Foundation,
Grant No.~0450696, and NSF-2013199; The Grainger Foundation;
Astrophysics Centre for Multi-messenger studies in Europe (ACME) EU
Grant No 101131928; and UNESCO.
\end{sloppypar}

%% file: main.bbl
\begin{thebibliography}{99}

\bibitem{CosmicRay1}B. Rossi, \textit{Supplemento a la Ricerca Scientifica }. \href{https://scholar.google.com/scholar_lookup?author=B+Rossi&journal=Supplemento+a+la+Ricerca+Scientifica&volume=1&pages=579&publication_year=1934}{\textbf{1} (1934) 579.}

\bibitem{CosmicRay2}K. Schmeiser et al., \textit{Die harten Ultrastrahlschauer}, \href{https://onlinelibrary.wiley.com/doi/10.1002/andp.19384240119}{\textit{Annalen der Physik}, \textbf{424} (1938) 161.}

\bibitem{CosmicRay3}W. Kolhörster et al., \textit{Gekoppelte Höhenstrahlen}, \href{https://link.springer.com/article/10.1007/BF01773491}{\textit{Naturwissenschaften}, \textbf{26} (1938) 576.}

\bibitem{CosmicRay5}P. Auger et al., \textit{Extensive Cosmic-Ray Showers}, \href{https://journals.aps.org/rmp/abstract/10.1103/RevModPhys.11.288}{\textit{Rev. Mod. Phys.}, \textbf{11} (1939) 288. }

\bibitem{Auger1}\textsc{Pierre Auger} collaboration, \textit{ The Pierre Auger Cosmic Ray Observatory}, \href{https://www.sciencedirect.com/science/article/pii/S0168900215008086?via%3Dihub}{\textit{Nucl. Instrum. Meth. A} \textbf{798} (2015) 172.}




\bibitem{FDPaper}\textsc{Pierre Auger} collaboration, \textit{The fluorescence detector of the Pierre Auger Observatory}, \href{https://www.sciencedirect.com/science/article/abs/pii/S0168900210008727?via%3Dihub}{\textit{NIM A}, \textbf{620} (2010) 227.}

\bibitem{AERA}\textsc{Pierre Auger} collaboration, \textit{Antennas for the Detection of Radio Emission Pulses from Cosmic-Ray induced Air Showers at the Pierre Auger Observatory}, \href{https://iopscience.iop.org/article/10.1088/1748-0221/7/10/P10011}{\textit{J. Instrum.} \textbf{7} (2012) P10011.}

\bibitem{ref:RDICRC2023} J. Pawlowsky et al., \textit{Status and expected performance of the AugerPrime
Radio Detector}, \href{https://pdfs.semanticscholar.org/34dd/e61f9fcdb5e83cade26271d9b6a2511d45de.pdf}{\textbf{PoS(ICRC2023)344}. }

\bibitem{AERA_energy2}\textsc{Pierre Auger} collaboration, \textit{Measurement of the radiation energy in the radio signal of extensive air showers as a universal estimator of cosmic-ray energy}, \href{https://journals.aps.org/prl/abstract/10.1103/PhysRevLett.116.241101}{\textit{Phys. Rev. Lett.} \textbf{116}, (2016) 241101.}


\bibitem{AERA_energy} \textsc{Pierre Auger} collaboration, \textit{Energy Estimation of Cosmic Rays with the Engineering Radio Array of the Pierre Auger Observatory,} \href{https://journals.aps.org/prd/abstract/10.1103/PhysRevD.93.122005}{\textit{Phys. Rev. D} \textbf{93} (2016) 122005.}

\bibitem{XmaxRadio1}\textsc{Pierre Auger} collaboration, \textit{Radio Measurements of the Depths of Air Shower Maxima at the Pierre Auger Observatory}, \href{https://journals.aps.org/prd/abstract/10.1103/PhysRevD.109.022002}{\textit{Phys. Rev. D} \textbf{109} (2024) 022002.}

\bibitem{XmaxRadio2}\textsc{Pierre Auger} collaboration, \textit{Demonstrating Agreement between Radio and Fluorescence Measurements of the Depth of Maximum of Extensive Air Showers at the Pierre Auger Observatory}, \href{https://journals.aps.org/prl/abstract/10.1103/PhysRevLett.132.021001}{\textit{Phys. Rev. Lett.} \textbf{132} (2024) 021001.}

\bibitem{Calibration_LPDA}\textsc{Pierre Auger} collaboration, \textit{Calibration of the Logarithmic-Periodic Dipole Antenna (LPDA) Radio Stations at the Pierre Auger Observatory using an Octocopter}, \href{https://iopscience.iop.org/article/10.1088/1748-0221/12/10/T10005}{\textit{J. Instrum.} \textbf{12} (2017) T10005. }

\bibitem{lofar}
K. Mulrey et al., \textit{Calibration of the LOFAR low-band antennas using the Galaxy and a model of the signal chain},  \href{https://www.sciencedirect.com/science/article/abs/pii/S0927650518302810?via%3Dihub}{\textit{Astropart. Phys.}, \textbf{111} (2019) 1.}

\bibitem{PerformanceSD}Orazio Zapparrata et al., \textit{The Time Evolution of the Surface Detector of the Pierre Auger Observatory}, \href{https://pos.sissa.it/444/266/pdf}{\textbf{PoS(ICRC2023)266}.}

\bibitem{PerformanceFD} \textsc{Pierre Auger} collaboration, \textit{Spectral calibration of the fluorescence telescopes of the Pierre Auger Observatory}, \href{https://www.sciencedirect.com/science/article/abs/pii/S0927650517301767}{\textit{Astropart. Phys.} \textbf{95} (2017) 44.}


\bibitem{Stations}\textsc{Pierre Auger} collaboration, \textit{Antennas for the Detection of Radio Emission Pulses from Cosmic-Ray induced Air Showers at the Pierre Auger Observatory}, \href{https://iopscience.iop.org/article/10.1088/1748-0221/7/10/P10011}{\textit{J. Instrum.} \textbf{7} (2012) P10011.}

\bibitem{Temperature} Rühle, C. F., \textit{Entwicklung eines schnellen eingebetteten Systems zur Radiodetektion kosmischer Strahlung.}, \href{https://www.iap.kit.edu/tunka-rex/downloads/Hiller_Thesis_KIT_TunkaRex.pdf}{\textit{Karlsruher Institut für Technologie (KIT)} (2014).}



\bibitem{beacon}\textsc{Pierre Auger} collaboration, \textit{Nanosecond-level time synchronization of autonomous radio detector stations using a reference beacon and commercial airplanes}, \href{https://iopscience.iop.org/article/10.1088/1748-0221/11/01/P01018}{\textit{J. Instrum.}, \textbf{11} (2016) P01018.}

\bibitem{NEC2}G. J. Burke et al., \textit{The Numerical Electromagnetics Code (NEC) - a brief history}, \href{https://ieeexplore.ieee.org/document/1331976/citations#citations}{\textit{IEEE Antennas Propag. Soc. Int. Symp.} \textbf{3} (2004) 2871.}


\bibitem{lfmap}
E. Polisensky, \textit{LFmap: A Low Frequency Sky Map Generating Program}, \href{https://leo.phys.unm.edu/~lwa/memos/memo/lwa0111.pdf}{\textit{Naval Research Lab} (2007).}

\bibitem{GSM}A. O. Costa et al., \textit{A model of diffuse Galactic radio emission from 10 MHz to 100 GHz}, \href{https://academic.oup.com/mnras/article/388/1/247/1011818}{\textit{MNRAS} \textbf{388} (2008) 247.}



\bibitem{GSM16}H. Zheng et al., \textit{An improved model of diffuse galactic radio emission from 10 MHz to 5 THz}
\href{https://academic.oup.com/mnras/article/464/3/3486/2514571}{\textit{MNRAS} \textbf{464} (2016) 3486.}


\bibitem{LFSM}J. Dowell et al.,  \textit{The LWA1 Low Frequency Sky Survey}, \href{https://academic.oup.com/mnras/article/469/4/4537/3815529?login=false}{\textit{MNRAS} \textbf{469} (2017) 4537. }


\bibitem{GMOSS} M. S. Rao et al., \textit{GMOSS: All-sky model of spectral radio brightness based on physical components and associated radiative processes}, \href{https://iopscience.iop.org/article/10.3847/1538-3881/153/1/26}{\textit{AJ} \textbf{153} (2017) 26.}

\bibitem{SSM} Q. Huang et al., \textit{A high-resolution self-consistent whole sky foreground model, }\href{https://ui.adsabs.harvard.edu/abs/2019SCPMA..6289511H/abstract}{\textit{Sci. China Phys. Mech. Astron.}, \textbf{62} (2019) 989511.}

\bibitem{ULSA}Y. Cong et al., \textit{An Ultralong-wavelength Sky Model with Absorption Effect}, \href{https://iopscience.iop.org/article/10.3847/1538-4357/abf55c}{\textit{Astrophys. J.}, \textbf{914} (2021) 128.}

\bibitem{busken2022} Büsken, M. et al., \textit{Uncertainties of the 30–408 MHz Galactic emission as a calibration source for radio detectors in astroparticle physics}, 
\href{https://www.aanda.org/articles/aa/full_html/2023/11/aa45382-22/aa45382-22.html}{\textit{A}\&\textit{A}, \textbf{679} (2023) A50.}

\bibitem{Haslam}Haslam, C. G. T. et al., \textit{The 408 MHz all-sky survey}, \href{https://ui.adsabs.harvard.edu/abs/1982A%26AS...47....1H/abstract}{\textit{A}\&\textit{AS}, \textbf{47} (1982) 1.}

\bibitem{StudySolarARENA} D. C. dos Santos et al., \textit{Study of solar activity with AERA data}, \href{https://pos.sissa.it/470/032/pdf}{PoS(ARENA2024)032.}


\bibitem{AugerPrime} \textsc{Pierre Auger} collaboration, \textit{The Pierre Auger Observatory and its Upgrade},  \href{http://scirevfew.net/index.php/sciencereviews/issue/view/4}{\textit{Science Reviews}, \textbf{1} (2020) 4.}

\bibitem{RD-paper} J. H. H$\ddot{\rm{o}}$randel et al., \textit{Precision measurements of cosmic rays up to the highest energies with a large radio array at the Pierre Auger Observatory},  \href{https://www.epj-conferences.org/articles/epjconf/abs/2019/15/epjconf_uhecr18_06005/epjconf_uhecr18_06005.html}{\textit{EPJ Web Conf.} \textbf{210} (2019) 06005.}

\end{thebibliography}
